\documentclass[twocolumn,amsmath,amssymb,pre]{revtex4}

\usepackage{graphicx}
\usepackage{bm}

\usepackage[T2A]{fontenc}
\usepackage[cp1251]{inputenc}

\usepackage{amsfonts}

\begin{document}
\title{Perturbation theory for solitons of the Fokas--Lenells equation : Inverse
scattering transform approach}
\author{V. M. Lashkin}
\email{vlashkin62@gmail.com} \affiliation{$^1$Institute for
Nuclear Research, Pr. Nauki 47, Kyiv 03028, Ukraine}
\affiliation{$^2$Space Research Institute, Pr. Glushkova 40 k.4/1,
Kyiv 03028,  Ukraine}

\begin{abstract}
We present perturbation theory based on the inverse scattering
transform method for solitons described by an equation with the
inverse linear dispersion law $\omega\sim 1/k$, where $\omega$ is
the frequency and $k$ is the wave number, and cubic nonlinearity.
This equation, first suggested by Davydova and Lashkin for
describing dynamics of nonlinear short-wavelength ion-cyclotron
waves in plasmas and later known as the Fokas--Lenells equation,
arises from the first negative flow of the Kaup--Newell hierarchy.
Local and nonlocal integrals of motion, in particular the energy
and momentum of nonlinear ion-cyclotron waves, are explicitly
expressed in terms of the discrete (solitonic) and continuous
(radiative) scattering data. Evolution equations for the
scattering data in the presence of a perturbation are presented.
Spectral distributions in the wave number domain of the energy
emitted by the soliton in the presence of a perturbation are
calculated analytically for two cases: (i) linear damping that
corresponds to Landau damping of plasma waves, and (ii)
multiplicative noise which corresponds to thermodynamic
fluctuations of the external magnetic field (thermal noise) and/or
the presence of a weak plasma turbulence.

\end{abstract}

\maketitle

\section{\label{sec1} Introduction}

Nonlinear evolution equations are widely used as models to
describe many phenomena in various fields of physics. The
classical examples are well-known universal models in dispersive
nonlinear media, such as the Korteweg-de Vries (KdV) and nonlinear
Schr\"{o}dinger (NLS) equations etc.
\cite{Dodd_book,Zakharov_book}. A common feature of nonlinear
evolution equations is the presence of dispersion and nonlinearity
which in some cases can effectively balance each other and lead to
soliton formation. Note that, generally speaking, most nonlinear
equations encountered in practical applications and admitting
analytical soliton solutions (most often in the one-dimensional
case, but there are examples for the two-dimensional
\cite{Petviashvili_book} and even three-dimensional cases
\cite{Lashkin2017}) are not completely integrable in the sense
that they do not have $N$-soliton solutions describing the elastic
collisions between solitons. Of particular interest are completely
integrable equations that are found in real physical applications
and are of important practical interest. In plasma physics, for
example, such equations are the KdV and NLS equations, derivative
nonlinear Schr\"{o}dinger (DNLS) equation, and  the
two-dimensional Kadomtsev-Petviashvili equation.

Davydova and Lashkin \cite{Lashkin1991} suggested a nonlinear
equation governing the dynamics of short-wavelength ion-cyclotron
waves in plasmas (the Bernstein modes) \cite{Trievel86,Akhiezer},
which in the one-dimensional case in dimensionless variables has
the form
\begin{equation}
\label{main} u_{xt}-u-i\sigma|u|^{2}u_{x}=0,
\end{equation}
where $u(x,t)$ is the slowly varying complex envelope of the
electrostatic potential at the ion gyrofrequency, and $\sigma=\pm
1$. Authors of Ref. \cite{Lashkin1991} found bright one-soliton
solution of Eq. (\ref{main}) corresponding to vanishing boundary
conditions at infinity, and then the same authors (with Fishchuck)
presented solutions of Eq. (\ref{main}) in the form of bright
algebraic soliton, dark and anti-dark solitons for the
nonvanishing boundary conditions, as well as solutions in the form
of nonlinear periodic waves in elliptic Jacobi functions
\cite{Lashkin1994}. Since the paper \cite{Lashkin1991} is not
quite widely available in the literature, note that all results of
Ref. \cite{Lashkin1991} are included in Ref. \cite{Lashkin1994},
except for the analysis of the modulation instability of a
monochromatic wave of finite amplitude, which is the simplest
solution of Eq. (\ref{main}). As was noted in Ref.
\cite{Lashkin1994}, the properties of the solitons of Eq.
(\ref{main}) differ from the properties of the solitons of the
KdV, NLS and DNLS equations. For example, the bright soliton of
Eq. (\ref{main}) can not be motionless, as well as for the KdV
soliton (unlike the  NLS and DNLS solitons), but, on the other
hand, its velocity does not depend on the soliton amplitude and is
an independent parameter unlike the KdV soliton (and like the
bright NLS and DNLS solitons). Equation (\ref{main}) is universal
in the sense that it contains only three terms of the second of
which corresponds to weak dispersion ($\omega\sim 1/k\ll 1$) and
the third to weak (cubic) nonlinearity. Here, $\omega$  and $k$
are the frequency and wave number respectively, where in the
linear part $u\sim \exp (i\omega t-ikx)$. The same situation holds
for well-known integrable models like the NLS and DNLS equations
with the weak dispersion $\omega\sim k^{2}\ll 1$ (and cubic
nonlinearity), and KdV equation with $\omega\sim k^{3}\ll 1$ (and
quadratic nonlinearity). Note that the weak dispersion and
nonlinearity in all cases follow from the physical derivation of
the corresponding equations. Later on, Fokas and Lenells showed
\cite{Fokas1995,Lenells2009_Nonlinearity} that Eq. (\ref{main}) is
completely integrable and corresponds to the first negative flow
of the Kaup--Newell hierarchy of the DNLS equation
\cite{Kaup1978}, and, therefore, can be solved by the inverse
transform scattering (IST) method \cite{Zakharov_book}. The
original version of the equation considered in Ref.
\cite{Fokas1995,Lenells2009_Nonlinearity} has been derived as an
integrable generalization of the NLS equation using bi-Hamiltonian
methods \cite{Fokas1995} and then as a model for nonlinear pulse
propagation in monomode optical fibers when certain higher-order
nonlinear effects are taken into account
\cite{Lenells2009_derivation}. Under this, the corresponding
equation in dimensionless variables is
\begin{equation}
\label{Fokas_Lenells1} iu_{t}-\nu u_{xt}+\gamma u_{xx}+\sigma
|u|^{2} (u+i\nu u_{x})=0,
\end{equation}
where $\nu$ and $\gamma$ are real constants, $\sigma\pm 1$ and by
gauge transformation and a change of variables can be reduced to
Eq. (\ref{main}). Lenells  rediscovered
\cite{Lenells2009_derivation} the bright one-soliton solution of
Ref. \cite{Lashkin1991} without using IST method. Bright
$N$-soliton solutions of Eq. (\ref{Fokas_Lenells1}) were obtained
by Lenells in Ref. \cite{Lenells2010_N-soliton} with the dressing
method  and for Eq. (\ref{main}) by the Hirota bilinearization
method in Ref. \cite{Matsuno_bright2012}. Dark $N$-soliton
solutions, which contain dark and anti-dark soliton solutions of
Ref. \cite{Lashkin1994}, were found by the bilinearization method
in Ref. \cite{Matsuno_dark2012}. In what follows, we will refer to
Eq. (\ref{main}) as the Davydova-Lashkin-Fokas-Lenells (DLFL)
equation.

In reality, in the DLFL equation which describes short-wavelength
nonlinear waves in a collisionless plasma, additional terms may be
present. They can include effects of dissipation due to finite
electric conductivity of plasma when taking into account the ion
viscosity, collisionless damping (linear and/or nonlinear Landau
damping), influence of external forces (external electric fields
under high-frequency plasma heating or particle beams),
inhomogeneity of the plasma density and/or the external magnetic
field, turbulent environment etc. \cite{Trievel86,Rukhadze84}.
These terms violate the integrability, but being small in many
important practical cases, they can be taken into account by
perturbation theory. The most powerful perturbative technique,
which fully uses the natural separation of the discrete and
continuous (i.e. solitonic and radiative) degrees of freedom of
the integrable equations, is based on the IST. Perturbation theory
based on the IST was first introduced by Kaup \cite{Kaup1976} and,
independently, by Karpman and Maslov
\cite{Karpman1977-1,Karpman1977}. An important contribution to the
development of IST perturbation theory has been made in Ref.
\cite{Newell1978}. A detailed review of the IST-based perturbation
theory for the KdV, focusing NLS, sine-Gordon and Landau-Lifshitz
equations is given by Kivshar and Malomed \cite{Kivshar1989}. For
the defocusing NLS equation with the nonvanishing boundary
conditions this was done in Refs. \cite{Lashkin_dark2004}, for the
modified NLS equation (combining the NLS and DNLS) in Ref.
~\cite{Doktorov1999} (for only discrete spectrum, using
Riemann-Gilbert problem) and in Ref.  ~\cite{Lashkin_MNLS2004}
(discrete and continuous spectrum), for the DNLS with the
vanishing and nonvanishing boundary conditions in Refs.
~\cite{Wyller1984,Lashkin_DNLS2006} and
Ref.~\cite{Lashkin_J_Phys2007} respectively.

Note that adiabatic equations for soliton parameters can be
obtained, generally speaking, without using the IST, and,
accordingly, even for non completely integrable equations. The
simplest techniques are based on integrals of motion of the
corresponding nonperturbed equation or use the variational
formalism and they have been applied successfully to various
problems in the theory of solitons.  However, these methods are
suitable for deriving the corresponding evolution equations only
in the lowest approximation, when an unperturbed instantaneous
shape of one soliton with slowly varying parameters is assumed.
They become inapplicable when considering the $N$-soliton solution
or when taking into account the effects that arise in higher
orders of perturbation theory. On the other hand, only the
perturbation theory based on the IST makes it possible to take
into account the excitation of continuous (radiative) degrees of
freedom, which leads to qualitatively new effects in one-soliton
dynamics. These effects include, in particular,
perturbation-induced emission of radiation by a soliton,
long-range corrections to the soliton's shape (tails), and the
generation of new (secondary) solitons
\cite{Kaup1976,Karpman1977-1,Karpman1977,Newell1978,Bullough1980,Kivshar1989,Lashkin_dark2004}.
In addition, the IST formalism allows one to obtain a criterion of
applicability of the adiabatic approach
\cite{Newell1978,Bullough1980,Kivshar1989,Lashkin_dark2004,Ablowitz_dark2011}.

The aim of this paper is to develop a perturbation theory based on
the IST for Eq. (\ref{main}) and to investigate bright soliton
propagation in the presence of a perturbation. The perturbed
equation is written in the form
\begin{equation}
\label{mainperturb} u_{xt}-u-i\sigma|u|^{2}u_{x}=p[u,u^{\ast}],
\end{equation}
where the perturbation is represented by the term $p[u,u^{\ast}]$.

The paper is organized as follows. In section ~\ref{sec1} we
review a theory of the  scattering transform for the linear
eigenvalue problem associated with the DLFL equation and calculate
the corresponding $N$-soliton Jost functions. In section
~\ref{sec2} integrals of motion are written in terms of the
discrete and continuous scattering data. Evolution equations for
the scattering data in the presence of perturbations are derived
in Sec. ~\ref{sec3}. As an application of the presented theory,
two cases: (i) linear damping that corresponds to Landau damping
of plasma waves, and (ii) multiplicative noise which corresponds
to fluctuations of the external magnetic field or the presence of
a weak plasma turbulence are considered in Sec. ~\ref{sec4} . The
conclusion is made in Sec. ~\ref{sec5}.

Regarding notations, we will use the stars for complex conjugation
(for matrices - elementwise). The Pauli matrices are
\begin{equation}
\sigma_{1}=
\begin{pmatrix}
0 & 1 \\
1 & 0
\end{pmatrix},\quad \sigma_{2}=
\begin{pmatrix}
0 & -i \\
i & 0
\end{pmatrix},\quad \sigma_{3}=
\begin{pmatrix}
1 & 0 \\
0 & -1
\end{pmatrix}.
\end{equation}

\section{\label{sec1} Inverse scattering transform for the DLFL equation}

Equation (\ref{main}) can be written as the compatibility
condition
\begin{equation}
\label{compatib} U_{t}-V_{x}+[U,V]=0 ,
\end{equation}
of two linear matrix equations \cite{Lenells2009_Nonlinearity} ,
where $M(x,t,\lambda)$ is a $2\times 2$ matrix-valued function and
$\lambda$ is a complex spectral parameter
\begin{gather}
\label{spec1} M_{x}=UM, \\
\label{spec2} M_{t}=VM,
\end{gather}
and
\begin{gather}
\label{U} U=-i\lambda^{2}\sigma_{3}+\lambda Q_{x}, \,\,\,\,
Q=\begin{pmatrix} 0 & u \\ \sigma u^{\ast} & 0
\end{pmatrix}
\\
\label{V}
V=\frac{i}{4\lambda^{2}}\sigma_{3}-\frac{i}{2\lambda}\sigma_{3}Q+\frac{i}{2}\sigma_{3}Q^{2},
\end{gather}
The Jost solutions $M^{\pm}(x,t,\lambda)$ of Eq. (\ref{spec1}) for
real $\lambda^{2}$ and for some fixed $t$ ($t$-dependence will be
omitted for now) are defined by the boundary conditions
\begin{equation}
\label{boundary} M^{\pm}(x,\lambda)\rightarrow E\equiv\exp
(-i\lambda^{2}\sigma_{3}x)
\end{equation}
as $x\rightarrow\pm \infty$. Since $\mathrm{Tr}\,U=0$, these
boundary conditions guarantee that $\mathrm{det}\,M^{\pm}=1$ for
all $x$. The matrix Jost solutions $M^{\pm}$ can be represented in
the form $M^{-}=(\varphi,-\tilde{\varphi})$ and
$M^{+}=(\tilde{\psi},\psi)$, where $\varphi$ and $\psi$ are
independent vector columns. The scattering matrix $S$
\begin{equation}
\renewcommand*{\arraystretch}{1.3}
S(\lambda)=
\begin{pmatrix}
 a(\lambda)  & -\tilde{b}(\lambda) \\
 b(\lambda) & \tilde{a}(\lambda)
\end{pmatrix},
\end{equation}
with $a\tilde{a}+b\tilde{b}=1$ relates the two fundamental
solutions $M^{-}$ and $M^{+}$
\begin{equation}
\label{Scat_matrix} M^{-}(x,\lambda)=M^{+}(x,\lambda)S(\lambda),
\end{equation}
so that
\begin{gather}
\varphi=a\tilde{\psi}+b\psi,
\label{co1} \\
\tilde{\varphi}=-\tilde{a}\psi+\tilde{b}\tilde{\psi}.
 \label{co2}
\end{gather}
It follows from Eqs. (\ref{spec1}) and (\ref{Scat_matrix}) that
matrices $M^{\pm}$ and $S$ have the parity symmetry properties,
\begin{equation}
\label{parity}
M^{\pm}(x,\lambda)=\sigma_{3}M^{\pm}(x,-\lambda)\sigma_{3}, \,\,
S(\lambda)=\sigma_{3}S(-\lambda)\sigma_{3},
\end{equation}
and the conjugation symmetry properties
\begin{alignat}{1}
\label{conjugation} M^{\pm}(x,\lambda)&=\sigma_{1}M^{\pm
\ast}(x,\lambda^{\ast})\sigma_{1}, \quad \,
\mathrm{if}\,\,\,\sigma =1,
\\
M^{\pm}(x,\lambda)&=\sigma_{2}M^{\pm
\ast}(x,\lambda^{\ast})\sigma_{2}, \quad \,
\mathrm{if}\,\,\,\sigma =-1 ,
\\
\label{conjugation1} \tilde{a}(\lambda)&=a^{\ast}(\lambda^{\ast}),
\quad \, \tilde{b}(\lambda)=-\sigma b^{\ast}(\lambda^{\ast}).
\end{alignat}
The coefficients $a(\lambda)$ and $b(\lambda)$ are
\begin{equation}
\label{S11} a(\lambda)=\mathrm{det}\,(\varphi,\psi), \,\,\,
b(\lambda)=\mathrm{det}\,(\tilde{\psi},\varphi).
\end{equation}
Taking into account the boundary conditions Eq. (\ref{boundary}),
the corresponding integral equations for $M^{\pm}$ can be obtained
from Eq. (\ref{spec1})
\begin{gather}
M^{\pm}(x,\lambda)=E(x,\lambda)-\lambda\int_{x}^{\pm\infty}E(x-y,\lambda)Q_{y}(y)
\nonumber \\
\times M^{\pm}(y,\lambda)\,dy. \label{int_pm}
\end{gather}
The standard analysis of these Volterra-type integral equations
yields the expressions for the Jost solutions at $\lambda=0$,
\begin{equation}
\label{asymp_lambda_0} M^{+}(x,0)=M^{-}(x,0)=E(x,0),
\end{equation}
and the asymptotics at $\lambda\rightarrow\infty$
\begin{gather}
\renewcommand*{\arraystretch}{1.4}
\psi(x,\lambda)\mathrm{e}^{-i\lambda^{2}x}=\mathrm{e}^{i\theta^{+}(x)}
\begin{pmatrix}
 \dfrac{u_{x}}{2i\lambda} \\ 1
\end{pmatrix} \left[1+O \left(\frac{1}{\lambda^{2}}\right)\right],\label{asymp_inf1}
 \\
\renewcommand*{\arraystretch}{1.4}
\varphi(x,\lambda)\mathrm{e}^{i\lambda^{2}x}=\mathrm{e}^{i\theta^{-}(x)}
\begin{pmatrix}
1 \\ \dfrac{\sigma u_{x}^{\ast}}{2i\lambda}
\end{pmatrix} \left[1+ O \left(
\frac{1}{\lambda^{2}}\right)\right], \label{asymp_inf2}
 \\
\renewcommand*{\arraystretch}{1.4}
\tilde{\psi}(x,\lambda)\mathrm{e}^{i\lambda^{2}x}=\mathrm{e}^{-i\theta^{+}(x)}
\begin{pmatrix}
1 \\ \dfrac{\sigma u_{x}^{\ast}}{2i\lambda}
\end{pmatrix} \left[1+ O \left(
\frac{1}{\lambda^{2}}\right)\right], \label{asymp_inf3}
 \\
\renewcommand*{\arraystretch}{1.4}
\tilde{\varphi}(x,\lambda)\mathrm{e}^{-i\lambda^{2}x}=-\mathrm{e}^{-i\theta^{-}(x)}
\begin{pmatrix}
 \dfrac{u_{x}}{2i\lambda} \\ 1
\end{pmatrix} \left[1+O \left(\frac{1}{\lambda^{2}}\right)\right],
\label{asymp_inf4}
\end{gather}
where we have introduced the notations
\begin{equation}
\label{theta_pm} \theta
^{\pm}(x)=\pm\frac{\sigma}{2}\int_{x}^{\pm\infty}|u_{y}|^{2}\,dy,
\end{equation}
From Eq. (\ref{asymp_inf1}) we have
\begin{equation}
\label{u_lim}
u_{x}=2i\lim_{\lambda\rightarrow\infty}\frac{\lambda\psi_{1}(\lambda)}{\psi_{2}(\lambda)}.
\end{equation}
The vector functions $\varphi(x,\lambda)$ and $\psi(x,\lambda)$
are analytically continuable to $\mathrm{Im}\,\lambda^{2}>0$,
while $\tilde{\varphi}(x,\lambda)$ and $\tilde{\psi}(x,\lambda)$
are analytically continuable to $\mathrm{Im}\,\lambda^{2}<0$. It
then follows from Eq. (\ref{S11}) that the coefficient
$a(\lambda)$ as a function of $\lambda$ is analytically
continuable to $\mathrm{Im}\,\lambda^{2}>0$ with the asymptotic at
$\lambda\rightarrow\infty$,
\begin{equation}
\label{S11_asymptotic} a(\lambda)=\exp (i\theta)(1+1/\lambda^{2}),
\end{equation}
where
\begin{equation}
\label{theta}
\theta=\theta^{-}+\theta^{+}=\frac{\sigma}{2}\int_{-\infty}^{\infty}|u_{x}|^{2}\,dx
.
\end{equation}
Likewise $\tilde{a}(\lambda)$ is analytically continuable to
$\mathrm{Im}\,\lambda^{2}<0$. For real $\lambda^{2}$ we have
$\lambda^{\ast}=\lambda$ if $\lambda^{2}>0$, and
$\lambda^{\ast}=-\lambda$ if $\lambda^{2}<0$, and then, using the
parity and conjugation properties Eqs. (\ref{parity}) and
(\ref{conjugation1}) one can write the normalization condition
$\mathrm{det}\,S(\lambda)=1$ as
\begin{equation}
|a(\lambda)|^{2}-\sigma\,\mathrm{sgn}\,\lambda^{2}|b(\lambda)|^{2}=1.
\end{equation}
The zeros $\lambda_{j}^{2}=\xi_{j}+i\eta_{j}$ ($j=1\dots N$) of
the function $a(\lambda)$ in the region of its analiticity
$\mathrm{Im}\,\lambda^{2}>0$ give the discrete spectrum of the
linear problem (\ref{spec1}) and correspond to solitons. In what
follows we will use the Kaup-Newell parametrization
\cite{Kaup1978} for the discrete eigenvalues $\lambda_{j}^{2}$
\begin{equation}
\lambda_{j}^{2}=\Delta^{2}_{j}(-\sigma\cos\gamma_{j}+i\sin\gamma_{j}),
\label{parametrizaion}
\end{equation}
where $\Delta_{j}>0$ and $0<\gamma_{j}<\pi$. With this
parametrization $\lambda_{j}$ and $-\lambda_{j}$ lie in the 1-st
and 3-rd quadrants respectively of the complex plane ($\pm
\lambda_{j}^{\ast}$ -- in the 2-st and 4-th quadrants
respectively). Under this, the functions $\varphi(x,\lambda_{j})$
and $\psi(x,\lambda_{j})$ are linearly dependent
\begin{equation}
\label{bj} \varphi(x,\lambda_{j})=b_{j}\psi(x,\lambda_{j}),\quad
\tilde{\varphi}(x,\lambda_{j}^{\ast})=-\sigma
b_{j}^{\ast}\tilde{\psi}(x, \lambda_{j}^{\ast}).
\end{equation}
The Jost coefficients $a(\lambda)$ and $b(\lambda)$ with real
$\lambda^{2}$ constitute the continuous spectrum scattering data,
and the set of complex numbers $\lambda_{j}$ and $b_{j}$ are the
discrete spectrum scattering data. One can express the function
$a(\lambda)$ for $\mathrm{Im\,\lambda^{2}>0}$ in terms of its
zeros and the values of $|b(\lambda)|$ on the contour
$\Gamma=\{\lambda;\,\mathrm{Im}(\lambda^{2})=0 \}$ (oriented as in
Fig. 1),
\begin{figure}
\includegraphics[width=2.5in]{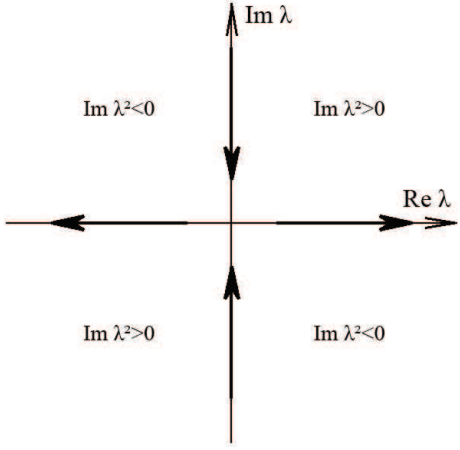}
\caption{\label{fig1}  The oriented integration contour $\Gamma$
and regions of analiticity of $a(\lambda)$ (the first and third
quadrants) and $\tilde{a}(\lambda)$ (the second and fourth
quadrants) respectively.}
\end{figure}
\begin{gather}
a(\lambda)=\prod_{j=1}^{N}\frac{\lambda^{2}-\lambda_{j}^{2}}
{\lambda^{2}-\lambda_{j}^{\ast 2 }}\exp \left\{i\theta
+\frac{1}{2\pi i} \right. \nonumber \\
\left.
\times\int_{\Gamma}\frac{\mu\ln(1+\sigma\,\mathrm{sgn}\,\mu^{2}|\,b(\mu)|^{\,2})}{\mu^{2}-
\lambda^{2}}\,d\mu\right\}. \label{mnls_reflection}
\end{gather}
From Eqs.  (\ref{S11}) and (\ref{asymp_lambda_0}) we have
$a(0)=1$, then, setting $\lambda=0$ in Eq.
(\ref{mnls_reflection}), one can find $\theta$ in terms of the
scattering data
\begin{equation}
\label{mnls_theta}
\theta=-4\sum_{j=1}^{N}\arg\lambda_{j}+\frac{1}{2\pi
}\int_{\Gamma}\frac{\ln(1+\sigma\,\mathrm{sgn}\,\mu^{2}|\,b(\mu)|^{2})}
{\mu}d\mu,
\end{equation}
where $\arg \lambda_{j}=\gamma_{j}/2$ for $\sigma=-1$ and $\arg
\lambda_{j}=\pi/2-\gamma_{j}/2$ for $\sigma=1$. From Eqs.
(\ref{mnls_reflection}) and (\ref{mnls_theta}) we have
\begin{gather}
a(\lambda)=\prod_{j=1}^{N}\frac{\lambda_{j}^{\ast 2
}}{\lambda_{j}^{2}}\frac{(\lambda^{2}-\lambda_{j}^{2})}
{(\lambda^{2}-\lambda_{j}^{\ast 2 })}\exp \left\{\frac{1}{2\pi i} \right. \nonumber \\
\left.
\times\int_{\Gamma}\frac{\lambda^{2}\ln(1+\sigma\,\mathrm{sgn}\,\mu^{2}|\,b(\mu)|^{\,2})}{\mu(\mu^{2}-
\lambda^{2})}\,d\mu\right\}. \label{mnls_reflection1}
\end{gather}
The  time evolution of the scattering data can be found, in a
standard way \cite{Zakharov_book}, from Eq. (\ref{V}) by
considering the limit $x\rightarrow\pm\infty$. Then
\begin{gather}
\label{dyn1} \lambda_{j}(t)=\lambda_{j}(0),
\\
\label{dyn2} b_{j}(t)=b_{j}(0)\exp[-i/(2\lambda^{2}_{j})t],
\\
\label{dyn3} b(\lambda,t)=b(\lambda,0)\exp[-i/(2\lambda^{2})t],
\end{gather}
and in the following we denote $\lambda_{j}(0)\equiv \lambda_{j}$,
$b_{j}(0)\equiv b_{j}$ and $b(\lambda,0)\equiv b(\lambda)$.

\begin{equation}
\label{S11_solit}
a(\lambda)=\prod_{j=1}^{N}\frac{\lambda_{j}^{\ast 2
}}{\lambda_{j}^{2}}\frac{(\lambda^{2}-\lambda_{j}^{2})}
{(\lambda^{2}-\lambda_{j}^{\ast 2 })}
\end{equation}
Since $S(\lambda)$  is diagonal in this case, it can be factorized
in such a way $S^{\,-}(\lambda)=S^{\,+}(\lambda)S(\lambda)$, that
the Jost solution matrices $M^{\,\pm}$ is expressed through a
common matrix $A(x,t,\lambda)$
\begin{equation}
\label{mnls_MAS}
M^{\,\pm}(x,t,\lambda)=A(x,t,\lambda)S^{\,\pm}(\lambda),
\end{equation}
where
\begin{equation}
\label{mnls_S+}
S^{\,+}=\mathrm{diag}\,\left(\prod_{j=1}^{N}\frac{\lambda_{j}}
{\lambda_{j}^{\ast}(\lambda^{2}-\lambda_{j}^{2})},
\prod_{j=1}^{N}\frac{\lambda_{j}^{\ast}}{\lambda_{j}
(\lambda^{2}-\lambda_{j}^{\ast 2})}\right)
\end{equation}
and $S^{\,-}=\sigma_{1}S^{\,+}\sigma_{1}$. Since $A(\lambda)$ is
analytical in the $\lambda$ plane, it follows from Eqs.
(\ref{asymp_inf1})-(\ref{asymp_inf4}) and (\ref{mnls_MAS}) that
diagonal and off-diagonal elements of the matrix
$A(\lambda)E^{-1}(\lambda)$ are polynomials in $\lambda$ of
degrees $2N$ and $2N-1$ respectively, and one can write
\begin{gather}
A(x,t,\lambda)=
\begin{pmatrix}  A_{11}^{(0)} & 0 \\
0 & A_{22}^{(0)}\end{pmatrix} \exp (-i\lambda^{2}\sigma_{3}x)
\nonumber
\\ +\sum_{k=1}^{N}\lambda^{2k-1}
\begin{pmatrix} \lambda A_{11}^{(k)} & A_{12}^{(k)}\\
A_{21}^{(k)} & \lambda A_{22}^{(k)}\end{pmatrix}\exp
(-i\lambda^{2}\sigma_{3}x), \label{mnls_A}
\end{gather}
where $A^{(k)}_{mn}$ with $k=0\dots N$ and $m,n=1,2$ are still
unknown functions. Setting here $\lambda=0$, we readily get from
Eqs. (\ref{asymp_lambda_0}) and (\ref{mnls_MAS}) the expressions
for the functions $A_{11}^{(0)}(x,t)$ and $A_{22}^{(0)}(x,t)$
\begin{equation}
\label{A00} A_{11}^{(0)}=
A_{22}^{(0)}=-\prod_{k=1}^{N}|\lambda_{k}|^{2}.
\end{equation}
From (\ref{bj}) and the fact $S^{\pm}=S^{\mp}$, it follows that
the columns of $A(x,t,\lambda)$ satisfy the relations
\begin{gather}
\label{mnls_A1}
A_{1}(x,t,\lambda_{j})=b_{j}(t)A_{2}(x,t,\lambda_{j}),
\\
\label{mnls_A2} A_{2}(x,t,\lambda_{j}^{\ast})=\sigma
b_{j}^{\ast}(t)A_{1}(x,t,\lambda_{j}^{\ast}).
\end{gather}
Substituting Eqs. (\ref{mnls_A}) and (\ref{A00}) into Eqs.
(\ref{mnls_A1}) and (\ref{mnls_A2}), one can obtain the linear
system of equations for the functions $A_{11}^{(k)}(x,t)$ and
$A_{12}^{(k)}(x,t)$
\begin{gather}
\sum_{k=1}^{N}\lambda_{j}^{
2k}A_{11}^{(k)}-\sum_{k=1}^{N}\lambda_{j}^{ 2k-1}\,\,c_{j}
A_{12}^{(k)} =\prod_{k=1}^{N}|\lambda_{k}|^{2},
\\
\sum_{k=1}^{N}\lambda_{j}^{\ast
2k}A_{11}^{(k)}-\sum_{k=1}^{N}\lambda_{j}^{\ast 2k-1}\,\, \sigma
c_{j}^{\ast -1} A_{12}^{(k)} =\prod_{k=1}^{N}|\lambda_{k}|^{2},
\label{systems}
\end{gather}
where
\begin{equation}
c_{j}=b_{j}\exp \left(2i\lambda_{j}^{2}x-i/(2\lambda^{2}_{j})t
\right).
\end{equation}
The remaining functions $A_{21}^{(k)}(x,t)$ and
$A_{22}^{(k)}(x,t)$ can be found from the symmetry properties Eq.
(\ref{conjugation})
\begin{equation}
A_{22}^{(k)}(x,t)=A_{11}^{(k) \ast}(x,t), \quad
A_{21}^{(k)}(x,t)=\sigma A_{12}^{(k) \ast}(x,t).
\end{equation}
Equations (\ref{mnls_MAS}) and (\ref{mnls_A}) determine the
$N$-soliton Jost solutions. By direct substitution one can check
that Eq. (\ref{mnls_A}) is compatible with Eqs. (\ref{spec1}),
(\ref{spec2}) and (\ref{mnls_MAS}) if and only if
\begin{equation}
\label{N-soliton_old} u_{x}=2i A_{12}^{(N)}/A_{22}^{(N)}.
\end{equation}
Equation (\ref{N-soliton_old}) can also be obtained from Eqs.
(\ref{u_lim}) and (\ref{mnls_MAS}).  Although in the case $N=1$
the integration for obtaining $u$ can be performed explicitly, it
is practically impossible already for $N\geqslant 2$. However,
using the dressing method, Lennels \cite{Lenells2010_N-soliton}
obtained an explicit formula for the $N$-soliton solution of Eq.
(\ref{Fokas_Lenells1}) (later on, an analogues formula was
obtained by Matsuno \cite{Matsuno_bright2012} by using the Hirota
bilinearization method) which in our notations (if $\sigma=-1$)
has the form
\begin{equation}
\label{N-soliton}
u(x,t)=\sum_{k,j=1}^{N}c_{k}^{\ast}(K^{-1})_{kj},
\end{equation}
where the elements of the $N\times N$ matrix $K$ are
\begin{equation}
\label{K-matrix}
K_{jk}=\frac{\lambda_{j}\lambda_{k}^{\ast}}{\lambda_{k}^{\ast
2}-\lambda_{j}^{ 2}}(\lambda_{j}+ \lambda_{k}^{\ast}c_{j}
c_{k}^{\ast}).
\end{equation}
The case $N=1$ corresponds to the one-soliton solution and from
Eqs. (\ref{N-soliton}) and (\ref{K-matrix}) one can readily get
\begin{equation}
\label{one-soliton} u=\frac{(\lambda_{1}^{\ast 2
}-\lambda^{2}_{1})c^{\ast}_{1}}{|\lambda_{1}|^{2}\left(\lambda_{1}+\lambda_{1}^{\ast}|c_{1}|^{2}\right)}
\end{equation}
On the other hand, from Eqs. (\ref{systems}) and
(\ref{N-soliton_old}) we have ($\sigma=-1$)
\begin{equation}
\label{one-soliton_x} u_{x}=2i(\lambda_{1}^{2}-\lambda_{1}^{\ast
2})\frac{(\lambda_{1}^{\ast}+\lambda_{1}|c_{1}|^{2})c_{1}^{\ast}}{(\lambda_{1}+\lambda_{1}^{\ast}|c_{1}|^{2})^{2}}.
\end{equation}
Next, we parametrize the complex numbers $\lambda_{1}$ and $b_{1}$
in terms of four real parameters $\Delta>0$, $0<\gamma<\pi$,
$x_{0}$ (the initial position of the soliton) and $\phi_{0}$ (the
initial phase) as
\begin{gather}
\label{parametrize1}\lambda_{1}^{2}=\Delta^{2}(\cos\gamma
+i\sin\gamma),
\\
\label{parametrize2} b_{1}=\exp
(2\Delta^{2}x_{0}\sin\gamma+i\phi_{0}).
\end{gather}
Then the one-soliton solution Eq. (\ref{one-soliton}) takes the
form
\begin{equation}
\label{one-soliton1} u=\frac{\sin\gamma\exp (-i\Phi)}{i\Delta\cosh
(z+i\gamma/2)},
\end{equation}
where
\begin{equation}
\label{z} z=2\Delta^{2}\left
(x-x_{0}+\frac{t}{4\Delta^{4}}\right)\sin\gamma  ,
\end{equation}
and
\begin{equation} \label{Fi} \Phi=2\Delta^{2}
\left(x-\frac{t}{4\Delta^{4}}\right)\cos\gamma +\phi_{0}.
\end{equation}
From Eqs. (\ref{one-soliton_x}), (\ref{parametrize1}) and
(\ref{parametrize2}) we have
\begin{equation}
\label{one-soliton1_x} u_{x}=-\frac{2\Delta\sin\gamma\cosh
(z-i\gamma/2)\exp (-i\Phi)}{\cosh^{2} (z+i\gamma/2)}.
\end{equation}
After integration in Eq. (\ref{one-soliton1_x}) we recovered Eq.
(\ref{one-soliton1}). Under this, $c_{1}=\exp (-z+i\Phi)$. An
explicit expression for $u$ in terms of the soliton amplitude and
phase is
\begin{equation}
\label{one-soliton2} u=\frac{\sin\gamma\exp
\{-i\Phi-i\arctan[\tanh z\tan
(\gamma/2)]\}}{i\Delta\sqrt{\cosh^{2}z-\sin^{2}(\gamma/2)}}.
\end{equation}
Earlier this solution was obtained by Davydova and Lashkin
\cite{Lashkin1991,Lashkin1994} without using the IST. The soliton
velocity $v$, amplitude $A$ and the characteristic halfwidth of
the soliton $w$ are
\begin{equation}
\label{velocity} v=-\frac{1}{4\Delta^{4}},\quad
A=\frac{\sin\gamma}{\Delta}, \quad
w=\frac{1}{2\Delta^{2}\sin\gamma}.
\end{equation}
It is seen that the soliton can not be motionless, and it moves
only in the negative direction of $x$-axis. Equation (\ref{main})
admits also rational $N$-soliton solutions, i.e. solitons with
algebraic decay at infinity. These solutions arise from the
solitons with exponential decay in the limit
$\gamma_{j}\rightarrow \pi$. In the case $N=1$,  from
(\ref{one-soliton1}) one can obtain
\begin{equation}
\label{algebraic} u=\frac{2\exp (-i\Phi)}{i\Delta-4\Delta^{3}y},
\end{equation}
where $y=x-x_{0}+t/(4\Delta^{4})$. This algebraic soliton of Eq.
(\ref{main}) was first obtained in Ref.~\cite{Lashkin1994} and
then rediscovered in Ref.~\cite{Lenells2009_Nonlinearity}. In
terms of the amplitude and phase, Eq. (\ref{algebraic}) takes the
form
\begin{equation}
\label{algebraic1} u=\frac{2\exp
[-i\Phi+i\,\mathrm{arccot}\,(4\Delta^{2}y)]}
{\Delta\sqrt{1+16\Delta^{4}y^{2}}}.
\end{equation}
Taking the Fourier transform in Eq. (\ref{one-soliton1}),
\begin{equation}
\label{Fourier} u(\omega,q)=\int_{-\infty}^{\infty}u(x,t)\exp
(-i\omega t+iqx)\,dx,
\end{equation}
one can  find the one-soliton field in the spectral space as
\begin{equation}
\label{Fourier1} u (\omega,q)=\frac{2\pi A\exp (-\gamma
\varkappa/2+i\psi_{0})}{i\cosh (\pi \varkappa/2)}\delta
\left(\omega-qv-\frac{\cos\gamma}{\Delta^{2}}\right),
\end{equation}
where $\delta (x)$ is the Dirac delta function, and
\begin{equation}
\varkappa=\cot\gamma-\frac{q}{k_{0}}, \quad  k_{0}=\frac{1}{w},
\quad \psi_{0}=\phi_{0}-k_{0}x_{0}.
\end{equation}
The presence of the $\delta$-function reflects the fact that a
single soliton is a stationary structure, i. e. it moves with the
constant velocity $v$, and the term $\cos\gamma/\Delta^{2}$
corresponds to the nonlinear frequency shift. Thus, as is seen
from Eqs. (\ref{Fourier}) and (\ref{Fourier1}), the soliton can be
treated as a localized wavepacket of monochromatic waves with
self-consistent amplitudes and phases.

Following Ref.~\cite{Gerdzhikov80}, and using the fact that the
$x$-part of the Lax pair Eq. (\ref{spec1}) is simply related to
the $x$-part of the Lax pair of the DNLS equation by the
replacement $u\rightarrow u_{x}$, one can write $u_{x}(x,t)$  in
terms of the scattering data and squared eigenfunctions of Eq.
(\ref{spec1}) as
\begin{equation}
\label{field}
u_{x}(x,t)=-4i\sum_{j=1}^{N}(C_{j}\psi_{1,j}^{2}+C_{j}^{\ast}\tilde{\psi}_{1,j}^{2})
-\frac{1}{\pi}\int_{\Gamma}(r\psi_{1}^{2}+\tilde{r}\tilde{\psi}_{1}^{2})\,d\lambda,
\end{equation}
where $r(\lambda)=b(\lambda)/a(\lambda)$ is the reflection
coefficient,
$\tilde{r}(\lambda)=\tilde{b}(\lambda)/\tilde{a}(\lambda)=r^{\ast}(\lambda^{\ast})$,
$C_{j}=b_{j}/\dot{a}_{j}$ with
$\dot{a}_{j}=da/d\lambda|_{\lambda=\lambda_{j}}$. Here, the
contribution of the discrete spectrum ($\sum_{j=1}^{N}$) is
explicitly separated from that of the continuous one ($\int
\,d\lambda$). The first term in Eq.~(\ref{field}) is the soliton
contribution, while the second one corresponds to the radiative
part of the field. In the asymptotic $t\rightarrow\infty$, a
generic initial field $u_{x}(x,0)$ will reshape itself into a set
of $N$ solitons (if any) and continuous radiation (quasilinear
waves). The latter always disperses away and decays, while the
solitons will propagate as coherent units. In the linear limit,
Eq.~(\ref{main}) describes the linear waves with the dispersion
relation $\omega=1/k$ (taking $u(x,t)\sim \exp (i\omega t-ikx)$)
that corresponds to the short-wavelength ion-cyclotron (ion
Bernstein) waves in a plasma \cite{Trievel86,Akhiezer} (see
\ref{appendA}). On the other hand, from Eq. (\ref{int_pm}) in the
linear limit we have $\psi_{1}\rightarrow 0$ and
$\tilde{\psi}_{1}\rightarrow \exp (-i\lambda^{2}x)$ (note that
this situation takes place as well for the NLS equation
\cite{Bullough1980}) so that ($t$-dependence remains the same as
before) the function $\tilde{\psi}_{1}^{2}$ simply reduces to
$\exp [-2i\lambda^{2}x+it/(2\lambda^{2})]$ and
$\tilde{r}(\lambda)$ is just the linear Fourier transform of
$u_{x}(x,t)$. This reflects the general property of the IST (see,
for example, Ref.~\cite{Bullough1980}): in the linear limit it is
equivalent to the usual Fourier method. Then, considering the
radiative component as a superposition of free waves governed by
the linearized Eq. (\ref{main}), one can conclude that the
spectral parameter $\lambda$ is connected to the wave number of
the emitted quasilinear waves $k$ by the relation
\begin{equation}
\label{lambda_and_k} k=2\lambda^{2},
\end{equation}
where $\lambda^{2}$ is real. Under this, the second term in Eq.
(\ref{field}) that corresponds to the radiative part of the field
$u_{x,\,rad}$ can be written as
\begin{equation}
\label{field_rad}
u_{x,\,rad}(x,t)=-\frac{1}{\pi}\int_{-\infty}^{\infty}[r(k)\psi_{1}^{2}(k)+\tilde{r}(k)\tilde{\psi}_{1}^{2}(k)]\,dk.
\end{equation}

\section{\label{sec2} Hamiltonian structure and Integrals of motion}

In Refs. \cite{Lenells2009_Nonlinearity,Lenells2009_derivation} it
was shown that Eq. (\ref{main}) arises from the first negative
flow of the Kaup–-Newell hierarchy of the DNLS equation
\cite{Kaup1978}
\begin{equation}
\label{hamilt_recur}
\renewcommand*{\arraystretch}{1.4}
\begin{pmatrix}u_{x} \\ \sigma
u_{x}^{\ast}
\end{pmatrix}_{t}=\sigma_{1}\frac{\partial}{\partial x}\,\mathrm{grad}\,H_{n-1}=J_{2}\,\mathrm{grad}\,H_{n},
\end{equation}
where $\mathrm{grad}\equiv(\delta/\delta u ,\sigma\,\delta/\delta
u_{x}^{\ast})^{\mathrm{T}}$, the superscript $\mathrm{T}$ denotes
transposition, and the operator $J_{2}$ are determined by
\begin{equation}
\label{operators_J2}
\renewcommand*{\arraystretch}{1.4}
J_{2}=\begin{pmatrix} -uu_{x} & i+\sigma u_{x}u^{\ast} \\
-i+\sigma uu_{x}^{\ast} & -u^{\ast}u_{x}^{\ast}
\end{pmatrix}.
\end{equation}
Then, the infinite sequence of conservation laws $H_{n}$ are
constructed recursively from the relation
\begin{equation}
\label{hamilt_recur2}
\mathrm{grad}\,H_{n+1}=J_{2}^{-1}\sigma_{1}\frac{\partial}{\partial
x}\,\mathrm{grad}\,H_{n}.
\end{equation}
where the inverse of $J_{2}$ is given by
\begin{equation}
\label{operators_J2-1}
\renewcommand*{\arraystretch}{1.4}
J_{2}^{-1}=\begin{pmatrix} u^{\ast}u_{x}^{\ast} & i+\sigma uu_{x}^{\ast} \\
-i+\sigma u_{x}u^{\ast} & uu_{x}
\end{pmatrix}.
\end{equation}
Gerdzhikov et. al. showed \cite{Gerdzhikov80} that the DNLS
equation is completely integrable Hamiltonian system, and
calculated the corresponding action-angle variables (see also
Ref.~\cite{Sasaki1982}). They obtained also the local and nonlocal
conservation laws in terms of the spectral data. Following
Ref.~\cite{Gerdzhikov80}, and using the simple relation between
Eq. (\ref{spec1}) and the $x$-part of the Lax pair of the DNLS
equation, one can obtain an explicit expression for the local and
nonlocal conservation laws $I_{n}$ of Eq. (\ref{main}) in the form
\begin{gather}
\renewcommand*{\arraystretch}{1.4}
I_{n}=\frac{1}{4|n|}\left\{ i\int_{-\infty}^{\infty}(\sigma
u_{x}^{\ast},u_{x})\hat{K}^{n}
\begin{pmatrix}u_{x} \\ \sigma u_{x}^{\ast}  \end{pmatrix} \,dx \right.\nonumber \\
\left.+4\int_{-\infty}^{\infty}\int_{x}^{\infty}(\sigma
u_{y}^{\ast},-u_{y})\hat{K}^{n+1}
\renewcommand*{\arraystretch}{1.4}
\begin{pmatrix}u_{y} \\ \sigma u_{y}^{\ast}
\end{pmatrix}\,dy\,dx\right\},
\label{C_m_phys}
\end{gather}
where the operator $\hat{K}$ is determined by
\begin{equation}
\label{K-operator}
\renewcommand*{\arraystretch}{1.4}
\hat{K}=\frac{i}{2}\left[1+i\begin{pmatrix}u_{x} \\ \sigma
u_{x}^{\ast}  \end{pmatrix} (\sigma
u_{x}^{\ast},-u_{x})\right]\sigma_{3}\frac{\partial}{\partial x}.
\end{equation}
Under this, the functionals $I_{n}$ are the expansion coefficients
in $\ln a(\lambda)$
\begin{equation}
\label{expansion_motions} \ln
a(\lambda)=i\theta+\sum_{n=1}^{\infty}\frac{I_{n}}{\lambda^{2n}}=\sum_{n=0}^{\infty}I_{-n}\lambda^{2n},
\end{equation}
and from Eq. (\ref{mnls_reflection}) one can  get the so-called
trace formulae
\begin{gather}
I_{n}=-\frac{1}{|n|}\sum_{j=1}^{N}\left(\lambda_{j}^{2n}-\lambda_{j}^{\ast
2n}\right)\nonumber \\
+\frac{i \, \mathrm{sgn}\,
n}{2\pi}\int_{\Gamma}\mu^{2n-1}\ln(1+\sigma\,\mathrm{sgn}\,\mu^{2}|\,b(\mu)|^{2})\,d\mu.
\label{C_m_spec}
\end{gather}
From the physical point of view, the quantities $u$ and $-u_{x}$
correspond to the electric potential and electrical field
respectively \cite{Lashkin1991,Lashkin1994} (see Appendix
\ref{appendA}).Then, the electrical energy  $E$ and the momentum
$P$ are
\begin{equation}
\label{energy}
E=2\sigma\theta=\int_{-\infty}^{\infty}|u_{x}|^{2}\,dx,
\end{equation}
\begin{equation}
\label{momentum}
P=iI_{-1}=\frac{i}{2}\int_{-\infty}^{\infty}(u^{\ast}u_{x}-uu_{x}^{\ast})\,dx.
\end{equation}
The energy and momentum can be explicitly expressed in terms of
the discrete (solitonic) and continuous (radiative) scattering
data. The expression for the energy Eq. (\ref{energy}) follows
from Eq. (\ref{mnls_theta}) where, for definiteness, we take
$\sigma=-1$,
\begin{equation}
\label{energy_spec1} E=4\sum_{j=1}^{N}\gamma_{j}
-\frac{1}{\pi}\int_{\Gamma}\frac{\ln(1-\mathrm{sgn}\,\lambda^{2}|\,b(\lambda)|^{2})}
{\lambda}\,d\lambda .
\end{equation}
The expression for the momentum $P$ follows from Eqs.
(\ref{C_m_spec}) and (\ref{momentum}) and has the form
\begin{equation}
\label{momentum_spec}
P=2\sum_{j=1}^{N}\frac{\sin\gamma_{j}}{\Delta_{j}^{2}}
-\frac{1}{2\pi}\int_{\Gamma}\frac{\ln(1-\mathrm{sgn}\,\lambda^{2}|\,b(\lambda)|^{2})}
{\lambda^{3}}\,d\lambda .
\end{equation}
Using the relation Eq. (\ref{lambda_and_k}) between the spectral
parameter $\lambda$ and the wave number $k$ of the emitted
quasilinear waves, one can also write for the energy and the
momentum
\begin{equation}
\label{energy_spec1_k} E=4\sum_{j=1}^{N}\gamma_{j}
+\int_{-\infty}^{\infty}\mathcal{E}_{rad}(k)\,dk,
\end{equation}
and
\begin{equation} \label{momentum_spec1_k}
P=4\sum_{j=1}^{N}\gamma_{j}
+\int_{-\infty}^{\infty}\mathcal{P}_{rad}(k)\,dk,
\end{equation}
where $\mathcal{E}_{rad}(k)$ and $\mathcal{P}_{rad}(k)$ are the
spectral energy and momentum densities (in the wave number domain)
carried by radiation respectively , determined by
\begin{equation}
\label{energy_spec1_k_density}
\mathcal{E}_{rad}(k)=-\frac{\ln(1-\mathrm{sgn}\,k|\,b(k)|^{2})}{2\pi
k}, \quad \mathcal{P}_{rad}(k)=\frac{\mathcal{E}_{rad}(k)}{k}.
\end{equation}
For a single soliton Eq. (\ref{one-soliton2}) one can express the
energy and the momentum through the soliton velocity $v$ and the
amplitude $A$ as
\begin{equation}
\label{energy_spec11_k} E=4\arcsin [(4|v|)^{-1/4}A]
+\int_{-\infty}^{\infty}\mathcal{E}_{rad}(k)\,dk,
\end{equation}
and
\begin{equation} \label{momentum_spec11_k}
P=2(4|v|)^{1/4}A +\int_{-\infty}^{\infty}\mathcal{P}_{rad}(k)\,dk.
\end{equation}

\section{\label{sec3} Dynamics of the scattering data in the presence of perturbations}

Equation (\ref{mainperturb}) can be cast in the matrix form
\begin{equation}
\label{cast}
\partial_{t}U-\partial_{x}V+[U,V]=P,
\end{equation}
where
\begin{equation}
 P=
\begin{pmatrix} 0 & \lambda p \\
\sigma\lambda  p^{\ast} & 0
\end{pmatrix}.
\end{equation}
From Eq. (\ref{cast}) and the fact that $M^{\pm}$ satisfies Eq.
(\ref{spec1}) one can get
\begin{equation}
\label{motion1} (\partial_{x}-U)(\partial_{t}-V)M^{\pm}=PM^{\pm}.
\end{equation}
Introducing a new unknown $J^{\pm}(x,t,\lambda)$ defined through
the relation
\begin{equation}
\label{Jpm} (\partial_{t}-V)M^{\pm}=M^{\pm}J^{\pm},
\end{equation}
and substitute Eq. (\ref{Jpm}) in Eq. (\ref{motion1}) one can see
that $J^{\pm}$ should satisfy the equation
$\partial_{x}J^{\pm}=M^{\pm\,-1}PM^{\pm}$, and, therefore, by
integrating we get
$J^{\pm}=C^{\,\pm}+\int_{\pm\infty}^{x}M^{\pm\,-1} PM^{\pm} dx'$,
where the constant matrices $C^{\,\pm}$ are determined from the
boundary conditions at $x\rightarrow\pm\infty$. Since
$V=i/(4\lambda^{2})\sigma_{3}$ as $x\rightarrow \pm\infty$, we
have from Eq. (\ref{Jpm}) $C^{\pm}=-i/(4\lambda^{2})\sigma_{3}$,
and, hence, the following equations of motion for $M^{\pm}$
\begin{equation}
 (\partial_{t}-V)M^{\pm}=M^{\pm}
\left[-\frac{i}{4\lambda^{2}}\sigma_{3}
+\int_{\pm\infty}^{x}(M^{\pm})^{-1} PM^{\pm} dx'\right].
\label{mo}
\end{equation}
Equation (\ref{mo}) is valid only for
$\mathrm{Im}\,\lambda^{2}=0$. Introducing the matrix
$M(x,t,\lambda)=(M_{1}^{-},M_{2}^{+})$, columns of which admit
analytical continuation to $\mathrm{Im}\,\lambda^{2}>0$, and
defining the  matrix $J(x,t,\lambda)=(J_{1}^{-},J_{2}^{+})$ we get
\begin{equation}\label{mot2}
(\partial_{t}-V)M=M J,
\end{equation}
one can  similarly obtain
\begin{gather}
\label{J1} \renewcommand*{\arraystretch}{1.4}
J_{1}=\begin{pmatrix} -\dfrac{i}{4\lambda^{2}} \\ 0
\end{pmatrix} +\int_{-\infty}^{x}M^{-1}P\varphi dx',
\\
\label{J2} \renewcommand*{\arraystretch}{1.4}
J_{2}=\begin{pmatrix} 0 \\ \dfrac{i}{4\lambda^{2}}
\end{pmatrix}
-\int_{x}^{\infty}M^{-1}P\psi dx'.
\end{gather}
Thus, we have the equations of motion valid for $\mathrm{Im
\,\lambda^{2}>0}$ except at $\lambda_{j}$, where $M$ fails to be
invertible. Making the assumption that the zeros
$\lambda=\lambda_{j}$ are simple, one can show (see below), that
each singularity is removable since $\det M=a(\lambda)$.
Differentiating Eq. (\ref{Scat_matrix}) with respect to $t$, and
using Eq. (\ref{mo}) yields
\begin{gather}
 \partial_{t}
S(t,\lambda)-\frac{i}{4\lambda^{2}}
[\sigma_{3},S(t,\lambda)]\nonumber \\
\label{motionS}
 =\int_{-\infty}^{\infty}(M^{+})^{-1}(x,t,\lambda)P
 M^{-}(x,t,\lambda)dx.
\end{gather}
The equations of motion for the coefficients $a(t,\lambda)$ and
$b(t,\lambda)$ are contained in Eq. (\ref{motionS}):
\begin{gather}
\label{eqa} \frac{\partial a}{\partial
t}=\lambda\int_{-\infty}^{\infty} (p\psi_{2}\varphi_{2}- \sigma
p^{\ast}\psi_{1}\varphi_{1})\,dx,
\\
\label{eqb} \frac{\partial b}{\partial
t}+\frac{i}{2\lambda^{2}}b=-\lambda\int_{-\infty}^{\infty}
(p\tilde{\psi_{2}}\varphi_{2}- \sigma
p^{\ast}\tilde{\psi_{1}}\varphi_{1})\,dx.
\end{gather}
The expression defining the zeros $\lambda_{j}(t)$ of
$a(t,\lambda)$ is $a(t,\lambda_{j}(t))=0$. Differentiating with
respect to $t$ gives
\begin{equation}
\label{ta}
\partial_{t}a(t,\lambda_{j}(t))+\frac{\partial\lambda_{j}}{\partial t}\,
\dot{a}_{j}=0,
\end{equation}
where $\dot{a}_{j}=da/d\lambda|_{\lambda=\lambda_{j}}$. Using
(\ref{eqa}) and (\ref{ta}) we have
\begin{equation}
\label{zeta} \frac{\partial\lambda_{j}^{2}}{\partial t}=-
\frac{2\lambda_{j}^{2}}{\dot{a}_{j}} \int_{-\infty}^{\infty}
(p\psi_{2,j}\varphi_{2,j}- \sigma
p^{\ast}\psi_{1,j}\varphi_{1,j})\,dx,
\end{equation}
where $\psi_{2,j}$, $\varphi_{2,j}$, $\psi_{1,j}$, and $
\varphi_{1,j}$ are the corresponding Jost solutions evaluated at
$\lambda=\lambda_{j}$. To obtain evolution equation for $b_{j}$,
we differentiate Eq. (\ref{bj}) with respect to $t$, use Eqs.
(\ref{mot2}), (\ref{J1}) and (\ref{J2}), and take the limit
$\lambda\rightarrow\lambda_{j}$ applying (since $\det
M(\lambda_{j})=a(\lambda_{j})=0$) the l'Hopitale rule and using
again Eq. (\ref{bj}). As a result, one obtains
\begin{gather}
 \frac{\partial b_{j}}{\partial
t}+\frac{i}{2\lambda_{j}^{2}}b_{j}=\frac{\lambda_{j}}{\dot{a}_{j}}\int_{-\infty}^{\infty}\left\{
p\,\varphi_{2}\frac{\partial}{\partial\lambda}\left(\varphi_{2}-b_{j}\psi_{2}\right)\right.
\nonumber \\
\left. -\sigma
p^{\ast}\,\varphi_{1}\frac{\partial}{\partial\lambda}
\left(\varphi_{1}-b_{j}\psi_{1}\right)\right\}dx',
 \label{eqg}
\end{gather}
where, after differentiating, the integrand is evaluated at
$\lambda=\lambda_{j}$. Equations (\ref{eqa}), (\ref{eqb}),
(\ref{zeta}) and (\ref{eqg}) describe the evolution of the
scattering data.

If $p[u,u^{\ast}]$ is a small perturbation, one can substitute the
unperturbed $N$-soliton solutions $\psi$, $\tilde{\psi}$,
$\varphi$ and $\tilde{\varphi}$  into the right-hand side of Eqs.
(\ref{eqa}), (\ref{eqb}), (\ref{zeta}) and (\ref{eqg}). This
yields evolution equations for the scattering data in the lowest
approximation of perturbation theory. This procedure can be
iterated to yield higher orders of perturbation theory. The
appearing hierarchy of equations are applied to arbitrary number
of solitons and, in particular, describe nontrivial many-soliton
effects in the presence of perturbations.

\section{\label{sec4} Perturbations in the DLFL equation}

In reality, as mentioned above, the DLFL equation describing
nonlinear ion-cyclotron waves in a collisionless plasma may
contain additional terms which in some cases are treated as small
perturbations. Below we give a summary of some of the physical
mechanisms that are often encountered in applications to plasma
physics, and write out the corresponding terms of the perturbation
$p$, where $\epsilon$ is a real constant depending on the specific
model. In particular, these are (the first two will be further
explored in detail in this section):
\begin{itemize}
\item Linear damping in collisionless plasma, see Eq.
(\ref{collisionless_damp}).
\item Multiplicative noise, see Eqs. (\ref{noise}) and
(\ref{correlator}).
\item Linear damping in collisional plasma \cite{Rukhadze84,Volland1984},
\begin{equation}
\label{dissip} p=\epsilon u_{xxx},
\end{equation}
where the perturbation has a diffusive character, and $\epsilon$
depends on the plasma ion viscosity and/or resistivity. This type
of perturbation is important in weakly ionized plasmas, in
particular, in the plasma of Earth's ionosphere
\cite{Volland1984}.
\item
Nonlinear Landau damping \cite{Ichikawa1973,Scorich2010},
\begin{equation}
\label{pLand} p=\epsilon
u_{x}\,\mathcal{P}\int_{-\infty}^{\infty}\frac{|u(x',t)|^{2}}{x'-x}dx',
\end{equation}
where $\mathcal{P}$ is the principal value of the integral.  The
non-local perturbation term $p$ represents the effect of resonant
particles on the wave modulations. The coefficient $\epsilon$
depends on the velocity distributions of the particle species.
\item
External pump used in the method of plasma heating in the ion
cyclotron range of frequency in plasma magnetic confinement
devices \cite{Rukhadze84,Adam1987},
\begin{equation}
\label{pump} p=\epsilon \exp (i \omega_{0} t),
\end{equation}
where the pump frequency $\omega_{0}$ is usually close to the
ion-cyclotron frequency.
\item Density or/and temperature gradient (or arbitrary
inhomogeneity),
and also curvature of the external magnetic field
\cite{Petviashvili_book,Rukhadze84,Adam1987},
\begin{equation}
\label{gardient} p=\epsilon F(x)u_{x},
\end{equation}
where $F(x)$ is often a linear function of $x$.
\item Weak interaction with the low-frequency magnetosonic wave
\cite{Petviashvili_book},
\begin{equation}
\label{alfen1} p=\epsilon bu_{x},
\end{equation}
with
\begin{equation}
\label{alfen2} b_{tt}-\chi b_{xx}=(|u|^{2})_{xx},
\end{equation}
where $b$ is the  perturbation of magnetic field associated with
the magnetosonic wave, and $\chi$ is the dimensionless Alfv\'{e}n
speed.
\end{itemize}

In this section we study the effect of the perturbing term in the
right-hand side of Eq. (\ref{mainperturb}) on a single soliton
($N=1$) described by Eq. (\ref{one-soliton1}) and take
$\sigma=-1$. Two different cases will be considered: (i) a linear
damping and (ii) an external multiplicative noise. From the
physical point of view, the first one corresponds to collisionless
Landau damping. The second case corresponds to fluctuations of an
external magnetic field or the influence of a turbulent
environment.

Note, that in the first case, the damping is irreversible and the
energy (both the soliton and the radiative parts) disappears in
such a system. In the second case, the perturbation has
conservative character and the total energy is conserved.

For the case $N=1$, the evolution equation for the discrete
scattering data $\lambda_{1}$ follows from Eq. (\ref{zeta}),
\begin{equation}
\label{eq_lambda1} \frac{\partial\lambda_{1}^{2}}{\partial
t}=-\frac{2\lambda_{1}^{2}}{\dot{a}_{1}}\int_{-\infty}^{\infty}(p\psi_{2,1}\varphi_{2,1}
+ p^{\ast}\psi_{1,1}\varphi_{1,1})dx,
\end{equation}
where the unperturbed one-soliton Jost functions
$\varphi_{1,1}$,$\varphi_{2,1}$,$\psi_{1,1}$ and $\psi_{1,1}$ are
determined by Eqs. (\ref{phi1_s})-(\ref{psi2_s}). After
calculating $\dot{a}_{1}$ from Eq. (\ref{S11_1}),  using Eqs.
(\ref{z}) and (\ref{Fi}), and making change of variables
$z\rightarrow -z$ in the second term of the integrand in Eq.
(\ref{eq_lambda1}), this equation can be written in a simple form
\begin{equation}
\label{eq_lambda1_simp} \frac{\partial\lambda_{1}^{2}}{\partial
t}=-i\lambda_{1}^{3}\int_{-\infty}^{\infty}\frac{\mathrm{e}^{z}}{\left(\lambda_{1}\mathrm{e}^{-z}
+\lambda_{1}^{\ast}\mathrm{e}^{z}\right)^{2}}[R(z)+R^{\ast}(-z)]dz,
\end{equation}
where $R(z)=p\exp (i\Phi)$. Taking into account Eqs.~(\ref{co1})
and (\ref{co2}), the evolution equation Eq.~(\ref{eqb}) for the
continuous scattering data $b(\lambda)$ can be rewritten as
\begin{gather}
\partial_{t}b+\frac{i}{2\lambda^{2}}b=-\frac{\lambda}{a(\lambda)}\int
[p(\tilde{a}\varphi_{2}+b\tilde{\varphi}_{2})\varphi_{2} \nonumber
\\
+p^{\ast}(\tilde{a}\varphi_{1}+b\tilde{\varphi}_{1})\varphi_{1}]\,dx,
\label{eqb11}
\end{gather}
and has the form
\begin{equation}
\label{eqb1}
\partial_{t}b+\frac{i}{2\lambda^{2}}b=-\frac{\lambda}{a(\lambda)}\int
(p\varphi_{2}^{2}+p^{\ast}\varphi_{1}^{2})\,dx,
\end{equation}
where we have used that $\varphi=a(\lambda)\tilde{\psi}$ for the
unperturbed scattering data $a(\lambda)$ and soliton Jost
functions $\varphi_{1,2}$ which are determined by
Eqs.~(\ref{S11_1}) and (\ref{phi1}).

\subsection{Linear damping}

Linear damping of waves in a plasma occurs due to collisions
or/and, in collisionless plasma, due to collisionless Landau
damping. The first case is more typical for a weakly ionized
plasma and the damping usually has the character of a
diffusion-type dissipation with the damping rate
$\Gamma_{k}=\mathrm{Im}\,\omega_{k}\sim k^{2}$, where $\omega_{k}$
is the frequency of plasma wave and $k$ is the wave number. The
second case corresponds to a fully ionized plasma where
collisionless damping dominates. In this paper, we restrict
ourselves to the case of collisionless plasma. As is well known,
the Landau damping rate $\Gamma_{k}$ for all types of plasma waves
is not a polynomial in the wavenumber $k$ and is an integral
operator in $x$-space \cite{Trievel86,Rukhadze84}, but in
practical applications (for optimal wavenumbers) it can be
estimated as independent of $k$ so that we use
$\Gamma_{k}=\Gamma\ll \mathrm{Re}\,\omega_{k}$. In physical
variables, the Landau damping rate $\Gamma$ for the considered
short-wavelength ion-cyclotron waves  can be found in Ref.
\cite{Perkins1976}. Thus, the perturbation term in Eq.
(\ref{mainperturb}) can be written as
\begin{equation}
\label{collisionless_damp}
 p=-\Gamma u_{x},
\end{equation}
and treated as a small perturbation.

\subsubsection{adiabatic approximation}

Substituting Eq. (\ref{collisionless_damp}) into Eq.
(\ref{eq_lambda1_simp}) and integrating, one can get
\begin{equation}
\label{dt_lambda1} \frac{\partial \lambda_{1}^{2}}{\partial
t}=2\Gamma\lambda_{1}^{2} (1-i\gamma-\gamma\cot\gamma).
\end{equation}
Separating the real and imaginary parts in Eq. (\ref{dt_lambda1})
we get equations for $\Delta$ and $\gamma$
\begin{gather}
\label{delta_t} \frac{\partial\Delta}{\partial t}=\Gamma\Delta
(1-\gamma\cot\gamma), \\ \label{gamma_t}
\frac{\partial\gamma}{\partial t}=-2\Gamma\gamma .
\end{gather}
From the latter equation we have $\gamma=\gamma_{0}\exp (-2\Gamma
t)$, where $\gamma_{0}$ is the initial value of $\gamma$ at the
moment $t=0$. Note that from Eq. (\ref{mainperturb}) and the
perturbation in the form Eq. (\ref{collisionless_damp}) one can
obtain the exact relation
\begin{equation}
\label{dt_theta} \frac{\partial \theta}{\partial
t}=-2\Gamma\theta,
\end{equation}
where $\theta$ is defined by Eq. (\ref{theta}) and is an integral
of motion in the absence of perturbations. In the terms of the
scattering data $\theta$ is determined by Eq. (\ref{mnls_theta})
and for the one-soliton solution Eq. (\ref{one-soliton1}) we have
$\theta=2\gamma$ so that we have Eq. (\ref{gamma_t}). Integrating
Eq. (\ref{delta_t}) yields
\begin{equation}
\label{Delta_zero} \Delta=\Delta_{0}\exp\left(\Gamma
t-\Gamma\int_{0}^{t}\gamma (t^{'})\cot\gamma (t^{'})
dt^{'}\right),
\end{equation}
where $\Delta_{0}$ is an initial value of $\Delta$ at $t=0$, and
after calculating the integral in Eq. (\ref{Delta_zero}) we find
$\Delta=\Delta_{0}$. Thus, as is  seen from Eq. (\ref{velocity}) ,
the soliton amplitude exponentially decays but the soliton
velocity remains constant. Note that for this type of
perturbation, the same dependence of the soliton parameters takes
place for the NLS equation \cite{Kaup1976}. We would like to
stress once again that the equations in the adiabatic
approximation can be obtained from the corresponding integrals of
motion without using the IST method.

\subsubsection{radiative effects}

The adiabatic approximation implies that $b(\lambda)=0$ and an
unperturbed instantaneous shape of the soliton is assumed. Now we
consider the radiative effects which are described by the
continuous spectrum scattering data $a(\lambda)$ and $b(\lambda)$.
In the presence of a perturbation the soliton emits radiation.
Indeed, as the soliton's amplitude is decreasing, as we have seen,
it is slowly loosing energy. The total energy exponentially decays
and soliton part of this energy is being dissipated away, but part
of it is transferred to the quasilinear waves, or in other words,
leads to excitation of the continuous spectrum.

In the case when a  perturbation has the form
Eq.~(\ref{collisionless_damp}) one can simplify Eq. (\ref{eqb1})
by using the relation
\begin{equation}
\label{relations22} (\varphi_{1}\varphi_{2})_{x}=\lambda
u_{x}\varphi_{2}^{2}-\lambda  u_{x}^{\ast}\varphi_{1}^{2},
\end{equation}
which directly follows from Eqs. (\ref{spec1}) and (\ref{U}).
Then, taking into account that $u_{x}\rightarrow 0$ as
$x\rightarrow\pm\infty$,  one can obtain
\begin{equation}
\label{eqb2}
\partial_{t}b+i\omega (\lambda)b=\frac{2\Gamma\lambda}{a(\lambda)}\int
u_{x}\varphi_{2}^{2}\,dx,
\end{equation}
where $\omega (\lambda)= 1/(2\lambda^{2})$. After substituting
Eqs.~(\ref{one-soliton_x}) and (\ref{phi1}) into Eq.~(\ref{eqb2})
and calculating the integral in the right-hand side of Eq.
(\ref{eqb2}),  we find
\begin{equation}
\label{eqb3}
\partial_{t}b+i\omega (\lambda)b=\Gamma\mathrm{e}^{i\beta (\lambda,\gamma)t} F(\lambda,\gamma)
\end{equation}
where the functions $\beta (\lambda,\gamma)$ and
$F(\lambda,\gamma)$ are defined by
\begin{equation}
\beta
(\lambda,\gamma)=\frac{\lambda^{2}}{2\Delta^{4}}-\frac{\cos\gamma}{\Delta^{2}},
\end{equation}
and
\begin{equation}
F(\lambda,\gamma)=\frac{2\pi\Delta\lambda^{3}
\left(\mathrm{e}^{-\gamma\alpha/2}-
\mathrm{e}^{\gamma\alpha/2}\lambda^{2}/\Delta^{2}\right)}
{[(\lambda^{2}-\Delta^{2})^{2}+4\lambda^{2}\Delta^{2}\sin^{2}(\gamma/2)]\cosh
(\pi\alpha/2)},
\end{equation}
respectively, where
$\alpha=\cot\gamma-\lambda^{2}/(\Delta^{2}\sin\gamma)$ and the
time dependence of $\gamma$ is determined by Eq. (\ref{gamma_t}).
Integration of Eq. (\ref{eqb3}) with an initial condition $b(0)=0$
yields
\begin{equation}
\label{eqb4} b(t)=\Gamma\mathrm{e}^{-i\omega
(\lambda)t}\int_{0}^{t}\mathrm{e}^{i\omega (\lambda)t^{'}+i\beta
(\lambda,\gamma)t^{'}}F(\lambda,\gamma)\,dt^{'}.
\end{equation}
There are two characteristic times of linear processes in the
model -- the damping time $\sim 1/\Gamma$ of linear waves with the
dispersion $\omega_{k}=1/k$, and the dispersive time $\sim
1/\omega_{k}$ at which the packet of linear waves spreads out due
to dispersion. At times $t\ll 1/\Gamma$, i. e. when the soliton
has not yet completely decayed, one may simply put $\beta
(\lambda,\gamma)=\beta (\lambda,\gamma_{0})$ and
$F(\lambda,\gamma)=F(\lambda,\gamma_{0})$. Then
\begin{equation}
\label{eqb5} b(t)=\frac{i\Gamma F(\lambda,\gamma_{0})}{[\omega
(\lambda)+\beta (\lambda,\gamma_{0})]}\left(\mathrm{e}^{-i\omega
(\lambda)t}-\mathrm{e}^{i\beta (\lambda,\gamma_{0})t}\right).
\end{equation}
In what follows we use the relation Eq. (\ref{lambda_and_k})
between the wave number of the emitted quasilinear waves $k$ and
the spectral parameter $\lambda$ and introduce the inverse soliton
halfwidth $k_{0}=1/w$ from Eq. (\ref{velocity}). From Eq.
(\ref{eqb5}) for $|b(t)|^{2}$ we have
\begin{equation}
\label{b_square}
|b(t)|^{2}=\frac{4\sin^{2}(\Omega_{k}t/2)}{\Omega_{k}^{2}}\Gamma^{2}|F(\mu,\gamma_{0})|^{2},
\end{equation}
where the function $|F(\mu,\gamma_{0})|^{2}$ is defined by
\begin{widetext}
\begin{equation}
|F(\mu,\gamma_{0})|^{2}=\frac{4\pi^{2}\mu^{3}\sin^{3}\gamma_{0}
\left(\mathrm{e}^{-\gamma_{0}\alpha/2}-
\mathrm{e}^{\gamma_{0}\alpha/2}\mu\sin\gamma_{0}\right)^{2}}
{[(1-\mu\sin\gamma_{0})^{2}+4\mu\sin\gamma_{0}\sin^{2}(\gamma_{0}/2)]^{2}\cosh^{2}
(\pi \alpha/2)},
\end{equation}
\end{widetext}
with $\alpha=\cot\gamma_{0}-\mu$, where $\mu=k/k_{0}$ is the ratio
of the soliton halfwidth to the wavelength of the emitted waves,
and
\begin{equation}
\label{Omega_k} \Omega_{k}=\omega_{k}-\omega^{NL}_{k}, \quad
\omega_{k}=\frac{1}{k^{2}},\quad
\omega^{NL}_{k}=kv+\frac{\cos\gamma_{0}}{\Delta^{2}}.
\end{equation}
The frequency $\Omega_{k}$ includes both the linear $\omega_{k}$
and nonlinear frequency shift $\omega^{NL}_{k}$ (see Eq.
(\ref{Fourier1})).
\begin{figure}
\includegraphics[width=2.8in]{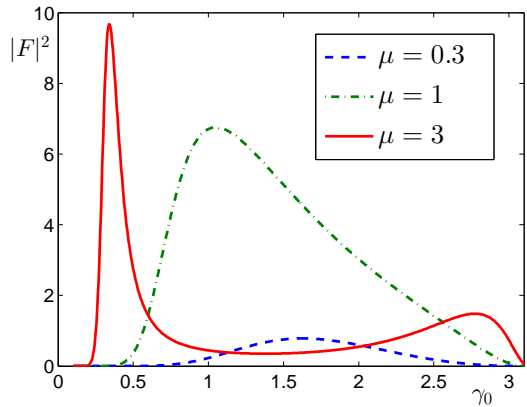}
\caption{\label{fig2} The function $|F(\mu,\gamma_{0})|^{2}$ for
different $\mu=k/k_{0}$.}
\end{figure}
The function $|F(\mu,\gamma_{0})|^{2}$ as a function of
$\gamma_{0}$ is plotted in Fig.~\ref{fig2} for different values
$\mu$. At times $t\gg 1/\Omega_{k}$ but still $\Gamma t\ll 1$, one
may consider an average over the period in Eq. (\ref{b_square})
and put $\overline{\mathstrut \sin^{2}}=1/2$, where the overbar
denotes the time average over the period. The emission intensity
is characterized by its power, i.e. the energy emission rate. The
absorbed emission power spectral density is
\begin{equation}
W_{rad}(k)=-\frac{d\mathcal{E}_{rad}(k)}{dt}=2\Gamma
\mathcal{E}_{rad}(k).
\end{equation}
If the perturbation is small enough then $|\,b(k)|\ll 1$ and from
Eq. (\ref{energy_spec1_k_density}) we have
\begin{equation}
\label{energy_spec1_k_density1}
\mathcal{E}_{rad}(k)=\frac{|\,b(k)|^{2}}{2\pi |k|},
\end{equation}
so that the absorbed emission power spectral density is
\begin{equation}
\label{power_spec_k}
W_{rad}(k)=\frac{2\Gamma^{3}|F(\mu,\gamma_{0})|^{2}}{\pi |k|
\Omega_{k}^{2}}.
\end{equation}
In particular, for $\gamma_{0}=\pi/2$ that corresponds to the
largest soliton amplitude for the fixed $\Delta$, we find
\begin{equation}
\label{power_spec_k1}
W_{rad}(\mu)=\frac{8\pi\Gamma^{3}k_{0}\mu^{4}\left(\mathrm{e}^{\pi\mu/4}
-\mu\mathrm{e}^{-\pi\mu/4}\right)^{2}}{(1+\mu^{2})^{4}\cosh^{2}
(\pi\mu/2)}.
\end{equation}
The corresponding dependence is shown in Fig.~\ref{fig3}
\begin{figure}
\includegraphics[width=2.8in]{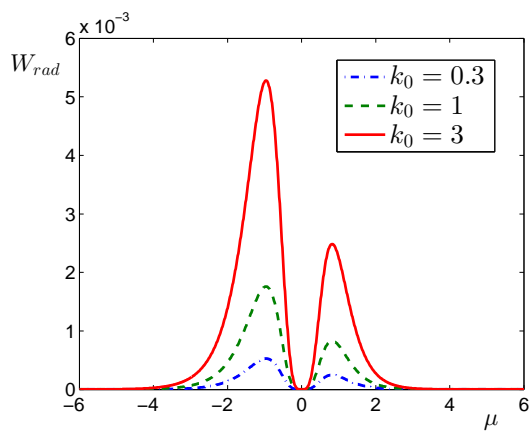}
\caption{\label{fig3} Spectral density of power emitted by the
soliton v.s. $\mu=k/k_{0}$ for different $k_{0}$, $\Gamma=0.1$.}
\end{figure}
In coordinate space, the continuous spectrum  appears as small
oscillations (quasilinear waves) moving away from the soliton,
where the wavelengths of these oscillations will be of the order
of the width of the soliton ($\mu\sim 1$) as is seen from
Fig.~\ref{fig3}. As the soliton decays and its width increases, so
does the typical wavelength of these oscillations. The width of
the soliton is of order $1/k_{0}$, which is also the
characteristic width $1/k$ of the emitted radiation.

At the end of this subsection, an important remark has to be made.
It is well known \cite{Newell1978,Bullough1980} that the action of
a perturbation in the form of linear damping (the effect of depth
change) on a soliton of the KdV equation leads to the appearance
of a long shelf containing as much mass as the original soliton.
Under this, the total motion is not adiabatic, for although the
soliton amplitude and the height the shelf itself are slowly
varying quantities, the range of the shelf is not. Thus, the
contribution of the continuous spectrum gives rise to a
qualitatively new effect. Kaup and Newell \cite{Newell1978} noted
that the presence of the shelf is connected with a singularity of
the reflection coefficient in the framework of the IST and then
correctly calculated the contribution corresponding to the
continuous spectrum (the shelf). The singularity of the reflection
coefficient on the real axis and, as a consequence, the emergence
of the shelf in the presence of a perturbation also takes place
for the dark soliton of the defocusing NLS equation
\cite{Lashkin_dark2004}. Note that the shelf of the dark soliton
of the defocusing NLS equation was studied in detail in
Ref.~\cite{Ablowitz_dark2011} using the so-called direct method.
In our case of an exponentially localized soliton Eq.
(\ref{one-soliton1}), one can see that a singularity in the
reflection coefficient $r(k)=b(k)/a(k)$ does not arise (just like
for the bright soliton of the focusing NLS equation) and the
adiabatic approximation is valid, but the singularity appears for
the algebraic soliton Eq. (\ref{algebraic}) since
$\gamma\rightarrow\pi$. Then the adiabatic approximation is
inapplicable and one can expect the emergence of a shelf.

\subsection{Multiplicative noise}

We now consider the perturbation in the form of a random
multiplicative noise. In the presence of fluctuations of the
magnetic field one can represent (for a given realization) the
magnetic field Eq. (\ref{B}) as $|u|^{2}\rightarrow
|u|^{2}+\varepsilon$, where $\varepsilon$ stands for the random
part of the field. Under this, the perturbation term in Eq.
(\ref{mainperturb}) takes the form
\begin{equation}
\label{noise} p=-i \varepsilon(x,t) u_{x},
\end{equation}
where $\varepsilon(x,t)$ is assumed to be real gaussian
homogeneous random field with the zero average $\langle
\varepsilon\rangle=0$ and the correlator
\begin{equation}
\label{correlator} \langle
\varepsilon(x,t)\varepsilon(x',t')\rangle=D(x-x')B(t-t'),
\end{equation}
where the angular brackets denote  ensemble averaging.

Here we would like to make an important remark. If the function
$\varepsilon(x,t)$ in Eq. (\ref{noise}) is independent of $x$,
then the DLFL equation (\ref{mainperturb}) with this kind of
perturbation $p$ turns out to be (somewhat unexpectedly)
completely integrable \cite{Kundu2010}. In this case, as can be
easily verified, in the Eq. (\ref{V}) for $V$ it is sufficient to
make the substitution $i/(4\lambda^{2})\rightarrow
i/(4\lambda^{2})-i\varepsilon (t)/2$ and the compatibility
condition Eq. (\ref{compatib}) still holds. In such a situation,
of course, the term in the right-hand side of Eq.
(\ref{mainperturb}) is not a perturbation. In the NLS, DNLS, DLFL
equations and some others with some specific additional terms, one
speaks of "integrable perturbations" \cite{KunduTMF}). Below we
consider only the case with the $x$-dependence in $\varepsilon
(x,t)$.

Note that there is an important difference between the
perturbations Eq. (\ref{collisionless_damp}) and Eq.
(\ref{noise}). In the case of a dissipative perturbation Eq.
(\ref{collisionless_damp}), the total energy vanishes over time,
while for the perturbation Eq. (\ref{noise}), as can be easily
shown, energy is conserved. Accordingly, for the dissipative
perturbation we considered the initial problem at time $t=0$. For
a conservative perturbation, we will focus on the stationary
regime and assume that the perturbation, which is absent at
infinity $t=-\infty$, turns on adiabatically. Thus, we use the
Fourier transform, and not the Laplace transform, for which the
problem is posed at $t=0$ and the causality condition requires
$u=0$ at $t<0$. Although the total energy is conserved, a
nonlinear interaction between soliton and radiation in the
presence of a perturbation results in energy redistribution
between the discrete and continuous parts of the spectrum.

One can see that in the approximation, when the right hand side in
Eq. (\ref{eq_lambda1}) depends only on unperturbed initial soliton
eigenvalue and unperturbed soliton Jost functions, we get
$\partial\langle\lambda_{1}^{2}/\partial t\rangle=0$ since
$\langle\varepsilon (x,t)\rangle=0$. We are interested in an
averaged emission power spectral density $\langle
W_{rad}(k)\rangle=d\langle E_{rad}(k)\rangle/dt$. In the follows
it is convenient to introduce the function $f=b\exp
(-i\omega_{k}t)$, and then assuming, as in the previous
subsection, that $|b(\lambda)|\ll 1$ and using Eq.
(\ref{energy_spec1_k_density1}) we have
\begin{equation}
\label{W_averaged} \langle
W_{rad}(k)\rangle=\frac{1}{\pi|k|}\mathrm{Re}\,\langle
f\partial_{t}f^{\ast}\rangle.
\end{equation}
Substituting Eq. (\ref{noise})  into Eq. (\ref{eqb1}) gives
\begin{equation}
\label{dt_f}
\partial_{t}f=\frac{i\lambda}{a(\lambda)}\mathrm{e}^{i\omega_{k}t}\int_{-\infty}^{\infty}\varepsilon
(x,t)G(x,t)\,dx,
\end{equation}
where $G(x,t)=u_{x}\varphi_{2}^{2}-u_{x}^{\ast}\varphi_{1}^{2}$.
Multiplying the right-hand side of Eq. (\ref{dt_f}) by $\exp
(\epsilon t)$ with an infinitely small $\epsilon>0$ that implies
adiabatically turning on a perturbation that was absent at
$t=-\infty$, and integrating, we get
\begin{equation}
\label{f}
f=\frac{i\lambda}{a(\lambda)}\int_{-\infty}^{t}\mathrm{e}^{i\omega_{k}t^{'}+\epsilon
t^{'}} \int_{-\infty}^{\infty}\varepsilon
(x,t^{'})G(x,t^{'})\,dxdt^{'}.
\end{equation}
Using Eq. (\ref{correlator}) and $|a(\lambda)|^{2}=1$ and Eq.
(\ref{lambda_and_k})
\begin{gather}
\langle
f\partial_{t}f^{\ast}\rangle=\frac{|k|}{2}\mathrm{e}^{-i\omega_{k}t}\int_{-\infty}^{t}
\int_{-\infty}^{\infty}\int_{-\infty}^{\infty}\mathrm{e}^{i\omega_{k}t^{'}} \nonumber \\
\times
B(t-t^{'})D(x^{'}-x^{''})G(x^{'},t^{'})G^{\ast}(x^{''},t)\,dt^{'}dx^{'}dx^{''}.
\end{gather}
Taking account that $B(t)$ and  $D(x)$ are even functions, we
write the Fourier transforms of $B(t)$, $D(x)$ and $G(x,t)$ in the
form
\begin{gather}
\hat{B}(\omega)=2\int_{0}^{\infty}\cos\omega t B(t)\,dt , \\
\hat{D}(q)=2\int_{0}^{\infty}\cos qx D(x)\,dx, \\
\hat{G}(q,t)=\int_{-\infty}^{\infty}\exp (-iqx)G(x,t)\,dx.
\label{Gq0}
\end{gather}
Then performing integration over $x^{'}$ and $x^{''}$ one can get
\begin{gather}
\langle
f\partial_{t}f^{\ast}\rangle=\frac{\lambda^{2}}{8\pi^{2}}\int_{-\infty}^{t}
\int_{0}^{\infty}\int_{-\infty}^{\infty}\hat{B}(\omega)\hat{D}(q)
\nonumber \\
\times \exp [i\omega_{k} (t^{'}-t)+\epsilon t^{'}]\cos
[\omega (t^{'}-t)] \nonumber \\
[\hat{G}(q,t^{'})\hat{G}^{\ast}(-q,t)+\hat{G}(-q,t^{'})\hat{G}^{\ast}(q,t)]\,dt^{'}d\omega
dq ,
\end{gather}
Using Eq. (\ref{relations22}) we have
$G(x,t)=(\varphi_{1}\varphi_{2})_{x}/\lambda$. Using Eq.
(\ref{Gq0}) we have for the vanishing at infinity boundary
conditions
\begin{equation}
\label{Gq}
\hat{G}(q,t)=\frac{iq}{\lambda}\int_{-\infty}^{\infty}\exp
(-iqx)\varphi_{1}\varphi_{2}\,dx.
\end{equation}
And then calculating integrals in Eq. (\ref{Gq}) we have
\begin{equation}
\label{Gq1}
\hat{G}(q,t)=\mathrm{e}^{-i(qv+\omega_{k}^{NL})t}\tilde{G}(q),
\end{equation}
where
\begin{equation}
\label{Gq2} \tilde{G}(q)=\frac{\pi q \mathrm{e}^{-2i\gamma}
\left(\mathrm{e}^{-\gamma\alpha/2}-
\mathrm{e}^{\gamma\alpha/2}\mu\sin\gamma\right)}{2\Delta [\exp
(-i\gamma)-\mu\sin\gamma]\cosh (\pi\alpha/2)},
\end{equation}
where $\alpha=\cot\gamma-\mu-q/k_{0}$, and $\omega_{k}^{NL}$ is
defined in Eq. (\ref{Omega_k}) and $\gamma=\gamma_{0}$ is the
initial value at $t=0$. Integrating over $t^{'}$ with Eq.
(\ref{Gq1}) and using Eq. (\ref{Gq2}) we find
\begin{gather}
\langle
f\partial_{t}f^{\ast}\rangle=\frac{|k|}{4i\Delta^{2}}\int_{-\infty}^{\infty}
\int_{-\infty}^{\infty}\frac{\hat{B}(\omega)\hat{D}(q)q^{2}}{(1+\mu^{2}\sin^{2}\gamma-\mu\sin
2\gamma)}
\nonumber \\
\times \frac{[\exp(-\gamma\beta)+
\mu^{2}\sin^{2}\gamma\exp(\gamma\beta)-2\mu\sin\gamma\cosh (\gamma
q/k_{0})]}
{[\cosh (\pi\beta)+\cosh (\pi q/k_{0})]} \nonumber \\
\times
\,\frac{\Omega_{k}(\Omega_{k}^{2}-\omega^{2}-q^{2}v^{2})}{[\Omega_{k}^{2}-(\omega+qv)^{2}]
[\Omega_{k}^{2}-(\omega-qv)^{2}]+i\epsilon}d\omega dq ,
\label{f_fc1}
\end{gather}
where $\beta=\cot\gamma - \mu$. The denominator in Eq.
(\ref{f_fc1}) has the simple poles at $\omega=\pm qv+\Omega_{k}$
and $\omega=\pm qv-\Omega_{k}$ on the real axis of the complex
$\omega$-plane and integration is performed according to the
prescription $\lim_{\epsilon\rightarrow
0}(y+i\epsilon)^{-1}=P(1/y)-i\pi\delta(y)$, where $P$ is the
symbol of the principal value. The pole on the real axis and the
appearance of the imaginary part means, as can be seen from Eqs.
(\ref{Omega_k}) and (\ref{f_fc1}), the resonance of a soliton with
quasilinear waves. A similar situation arises, for example, in the
linear theory of plasma when solving the Vlasov kinetic equation
in Fourier space results in the pole corresponding to the
wave-particle resonance leading to collisionless damping of the
wave at which the total energy of the wave-particle system is
conserved. Substituting Eq. (\ref{f_fc1}) into Eq.
(\ref{W_averaged}) one can find for the averaged emission power
spectral density
\begin{gather}
\langle W_{rad}(k)\rangle=-\frac{1}{4\Delta^{2}}
\int_{-\infty}^{\infty}\hat{D}(q)V_{k}(q) [\hat{B}(\Omega_{k}+qv)
\nonumber
\\
+\hat{B}(\Omega_{k}-qv) +\hat{B}(-\Omega_{k}+qv)
+\hat{B}(-\Omega_{k}-qv) ]\,dq , \label{f_fc2}
\end{gather}
where the function $V_{k}(q)$ is defined by
\begin{widetext}
\begin{equation}
V_{k}(q)=\frac{q^{2}[\exp(-\gamma\beta)+
\exp(\gamma\beta)\mu^{2}\sin^{2}\gamma-2\mu\sin\gamma\cosh (\gamma
q/k_{0})]} {(1+\mu^{2}\sin^{2}\gamma-\mu\sin 2\gamma)[\cosh
(\pi\beta)+\cosh (\pi q/k_{0})]}. \label{Vq}
\end{equation}
\end{widetext}
 As is seen from Eqs. (\ref{f_fc2}) and (\ref{Vq}),
if $\hat{D}(q)=\delta (q)$ that is the noise does not depend on
the spatial coordinate, we have $\langle W_{rad}(k)\rangle=0$.
This is in accordance with the fact that, as said above, the Eq.
(\ref{mainperturb}) with the perturbation Eq. (\ref{noise})
depending only on $t$ is completely integrable
\cite{Kundu2010,KunduTMF}. Consider now the case when the spatial
part $\varepsilon(x)$ of the random function $\varepsilon(x,t)$
has the form
\begin{equation}
\varepsilon(x)=\varepsilon_{0}\cos(q_{0}x+\vartheta),
\end{equation}
where the random amplitude $\varepsilon_{0}$ is a zero mean,
normally distributed value with variance $\sigma^{2}$, and the
random phase $\vartheta$ is uniformly distributed between $0$ and
$2\pi$. The correlation function of such a process is
$D(x)=(\sigma^{2}/2)\cos(q_{0}x)$ or, in the wave number domain
\begin{equation}
\label{peak}
\tilde{D}(q)=\frac{\sigma^{2}}{4}[\delta(q-q_{0})+\delta(q+q_{0})].
\end{equation}
In this case the space noise has an infinite correlation length
and is concentrated at the wave number $q_{0}$. Then the averaged
power spectral density $\langle W_{rad}(k)\rangle$ can be written
in a closed form for the arbitrary frequency correlator
$\tilde{B}(\omega)$, and, for example, for the gaussian shape
\begin{equation}
\hat{B}(\omega)=B_{0}\exp (-\omega^{2}/\omega_{c}^{2}),
\end{equation}
where $\tau_{c}=1/\omega_{c}$ is the correlation time, one can
write
\begin{gather}
\langle W_{rad}(k)\rangle=-2\sqrt{|v|}\sigma^{2}B_{0}V_{k}(q_{0})
\exp[-(\Omega_{k}^{2}+q^{2}_{0}v^{2})/\omega_{c}^{2}]
\nonumber \\
\times\cosh (2\Omega_{k}q_{0}v/\omega_{c}^{2}).
\end{gather}
The case $\tau_{c}=0$ is the $\delta$-time correlated field (the
white noise). Then consider two important cases when the frequency
correlator in Eq. (\ref{f_fc2}) has the form
\begin{equation}
\hat{B}(\omega)=B_{0}[\delta (\omega-\omega_{k})+\delta
(\omega-\omega_{k})].
\end{equation}
The first case corresponds to thermodynamic fluctuations of the
magnetic field (thermal noise) \cite{Sitenko} in Eq. (\ref{B}) and
the second one to the presence of weak electromagnetic turbulence
\cite{Tsytovich}. Then
\begin{gather}
\langle W_{rad}(k)\rangle=-\sqrt{|v|} \sum_{\pm}\hat{D}(\pm
q_{1})V_{k}(\pm q_{1})
 \nonumber \\
+\hat{D}(\pm q_{2})V_{k}(\pm q_{2}),
\end{gather}
where $q_{1}=\omega_{k}^{NL}/v$ and
$q_{2}=(\omega_{k}^{NL}-2\omega_{k})/v$. Note that in the first
case the correlator $\hat{D}(q)$ does not depend on $q$
\cite{Sitenko} and is simply proportional electron and ion
temperatures (in the case under consideration, the derivation of
the specific expression for the correlator is beyond the scope of
this paper). In the case of a weak turbulence, the correlator
$\hat{D}(q)$  usually has a Lorentz shape \cite{Tsytovich}. In
contrast to the damping case, the emitted energy is not damped but
transferred to infinity. Far from the emitting soliton, the
radiation field looks like a traveling monochromatic wave. Since
the averaged total energy is conserved, from Eq.
(\ref{energy_spec1_k}) one can immediately write for the averaged
soliton parameter $\gamma$
\begin{equation}
4\frac{\langle\gamma\rangle}{\partial t}=-\int_{-\infty}^{\infty}
\langle W_{rad}(k)\rangle \, dk.
\end{equation}

In conclusion of this subsection, we note that the special case of
a perturbation of the form $\varepsilon (x,t)=gx$ in Eq.
(\ref{noise}),  when $g$ is the deterministic constant,
corresponds to the gradient of the external magnetic field. Under
this, using the relation Eq. (\ref{relations22}), the
corresponding evolution equations for $\lambda_{1}$ and
$b(\lambda)$ are simplified and, in particular, it can be shown
that, depending on the sign of $g$, the amplitude of the soliton
either decreases or increases. Detailed analysis will be presented
elsewhere.

\section{\label{sec5} Conclusions}
We have presented a perturbation theory based on the IST for
solitons of the completely integrable equation with the inverse
linear dispersion law $\omega\sim 1/k$ and cubic nonlinearity.
This equation governs the dynamics of nonlinear short-wavelength
ion-cyclotron waves in plasmas. An approach based on the IST fully
uses the natural separation of the discrete and continuous degrees
of freedom of the unperturbed equation. Local and nonlocal
integrals of motion, in particular the energy and momentum of
nonlinear ion-cyclotron waves, were explicitly expressed in terms
of the discrete (solitonic) and continuous (radiative) scattering
data. We have derived evolution equations for the scattering data
in the presence of perturbations.  As an application, we
considered two cases: (i) linear damping that corresponds to
Landau damping of plasma waves, and (ii) multiplicative noise
which corresponds to thermodynamic fluctuations of the external
magnetic field (thermal noise) and/or the presence of a weak
turbulence. In both cases spectral distributions of the energy
emitted by the soliton  were calculated analytically. In the case
of the linear damping, the amplitude of the soliton decreases
exponentially while its velocity remains constant.

\appendix

\section{\label{appendA} physical application}

For a plasma in an uniform external magnetic field
$\mathbf{H}_{0}=H_{0}\hat{\mathbf{z}}$ oriented along the
$z$-axis, a general linear dispersion relation for the
electrostatic ion-cyclotron waves (the Bernstein modes) in the
short-wavelength limit $k_{\perp}\rho_{i}\gg 1$ under the
conditions $k_{\perp}\rho_{e}\ll 1$ and $\omega\ll k_{z}v_{Te}$
is,
\begin{gather}
\omega (\mathbf{k})=n\omega_{ci}\left[1+\frac{1}{\sqrt{2\pi}(1
+T_{i}/T_{e})k_{\perp}\rho_{i}}\right] \nonumber
\\
\equiv n\omega_{ci}[1+R(k_{\perp})], \label{disp_phys}
\end{gather}
where $R(k)_{\perp}\ll 1$ \cite{Akhiezer}. Here $\omega$ and
$\mathbf{k}$ are the frequency and wave vector respectively,
$\omega_{ci}$ is the ion gyrofrequency, $\rho_{\alpha}$,
$v_{T\alpha}$ and $T_{\alpha}$ are the Larmor radius, thermal
velocity and temperature of particle species $\alpha$ ($e$ for
electrons and $i$ for ions) respectively,  $n=1,2,\dots$ and next
only the case of the lowest harmonics $n=1$ is considered. The
nonlinear equation \cite{Lashkin1991,Lashkin1994} for the envelope
$\Phi$ of the electrostatic potential $\tilde{\Phi}$ at the ion
gyrofrequency
\begin{equation}
\label{envelope} \tilde{\Phi}=\frac{1}{2}[\Phi\exp
(-i\omega_{ci}t)+\mathrm{c}.\,\mathrm{c}]
\end{equation}
has the form,
\begin{equation}
\label{equation2D}
\Delta_{\perp}\left(\frac{i}{\omega_{ci}}\frac{\partial\Phi}{\partial
t}-\hat{R}\Phi\right)=\nabla_{\perp}\cdot (h\nabla\Phi),
\end{equation}
where $\Delta_{\perp}=\partial^{2}/\partial
x^{2}+\partial^{2}/\partial y^{2}$,  $\nabla_{\perp}
=(\partial/\partial x,\partial/\partial y)$, and the operator
$\hat{R}$ is defined by
\begin{equation}
\label{R} \hat{R}\Phi (\mathbf{r},t)=\int
R(k_{\perp})\hat{\Phi}(\mathbf{k}_{\perp},t)\exp
(i\mathbf{k}_{\perp}\cdot \mathbf{r}),
\end{equation}
and
\begin{equation}
\label{B} h=\frac{\delta
H_{z}}{H_{0}}=-\frac{\omega_{pe}^{2}mT_{i}|\Phi|^{2}}{4H_{0}^{2}T_{e}},
\end{equation}
where $\delta H_{z}$ is the nonlinear perturbation of the magnetic
field, $\omega_{pe}$ and $m$ are the electron plasma frequency and
the electron mass respectively. In the one-dimensional case, and
in the dimensionless variables
\begin{equation}
x\rightarrow \frac{x}{\sqrt{2\pi}(1+T_{i}/T_{e})\rho_{i}},\quad
u\rightarrow \Phi\frac{\omega_{pe}\sqrt{mT_{i}}}{2H_{0}T_{e}}
\end{equation}
equation (\ref{equation2D}) reduces to Eq. (\ref{main}).

\section{\label{appendB} one-soliton scattering data and Jost solutions}

One-soliton scattering data are
\begin{equation}
\label{S11_1} a(\lambda)=\frac{\lambda_{1}^{\ast
2}(\lambda^{2}-\lambda_{1}^{2})}
{\lambda_{1}^{2}(\lambda^{2}-\lambda_{1}^{\ast 2})},\quad
b(\lambda)=0, \quad (\mathrm{Im}\,\lambda^{2} \geqslant 0).
\end{equation}
One-soliton Jost solutions are
\begin{equation}
\label{phi1}
\renewcommand*{\arraystretch}{1.3}
\begin{pmatrix} \varphi_{1} \\ \varphi_{2} \end{pmatrix}
=\mathrm{e}^{-i\lambda^{2}x}\frac{\lambda_{1}^{\ast}}{\lambda_{1}
(\lambda^{2}-\lambda_{1}^{\ast 2})}\begin{pmatrix}
\lambda^{2}A_{11}^{(1)}-|\lambda_{1}|^{2} \\
\lambda A_{21}^{(1)}
\end{pmatrix},
\end{equation}
and
\begin{equation}
\label{psi1}
\renewcommand*{\arraystretch}{1.3}
\begin{pmatrix} \psi_{1} \\ \psi_{2} \end{pmatrix}
=\mathrm{e}^{i\lambda^{2}x}\frac{\lambda_{1}^{\ast}}{\lambda_{1}
(\lambda^{2}-\lambda_{1}^{\ast 2})}\begin{pmatrix}
 \lambda A_{12}^{(1)} \\
\lambda^{2}A_{22}^{(1)}-|\lambda_{1}|^{2}
\end{pmatrix} ,
\end{equation}
where
\begin{equation}
\label{A11}
A_{11}^{(1)}=\frac{\lambda_{1}^{\ast}-\lambda_{1}\sigma|c_{1}|^{2}}{\lambda_{1}
-\lambda_{1}^{\ast}\sigma|c_{1}|^{2}}, \, \, \,
A_{22}^{(1)}=A_{11}^{ (1)\ast},
\end{equation}
and
\begin{equation}
\label{A12} A_{12}^{(1)}=-\frac{(\lambda^{2}_{1}-\lambda_{1}^{\ast
2 })\sigma c_{1}^{\ast}}{\lambda_{1}
-\sigma\lambda_{1}^{\ast}|c_{1}|^{2}}, \, \, \,
A_{21}^{(1)}=\sigma A_{12}^{ (1)\ast} ,
\end{equation}
where $c_{1}=\exp (-z+i\Phi)$ and $z$ and $\Phi$ are determined by
Eq. (\ref{z}) and (\ref{Fi}) respectively. The remaining Jost
solutions can be found from the symmetry properties Eq.
(\ref{conjugation})
\begin{gather}
\tilde{\varphi}_{1}(\lambda)=\sigma\varphi_{2}^{\ast}(\lambda^{\ast}),\,\,
\,
\tilde{\varphi}_{2}(\lambda)=\varphi_{1}^{\ast}(\lambda^{\ast}),
\\
\tilde{\psi}_{1}(\lambda)=\psi_{2}^{\ast}(\lambda^{\ast}),\,\, \,
\tilde{\psi}_{2}(\lambda)=\sigma\psi_{1}^{\ast}(\lambda^{\ast}).
\end{gather}
One soliton Jost solutions evaluated at $\lambda_{1}$ are
\begin{gather}
\label{phi1_s}
\varphi_{1,1}(\lambda_{1})=-\mathrm{e}^{-i\lambda_{1}^{2}x}
\frac{\lambda_{1}^{\ast}\sigma|c_{1}|^{2}}{(\lambda_{1}-\lambda_{1}^{\ast}\sigma|c_{1}|^{2})},
\\
\label{phi2_s}
\varphi_{2,1}(\lambda_{1})=\mathrm{e}^{-i\lambda_{1}^{2}x}
\frac{\lambda_{1}^{\ast}c_{1}}{(\lambda_{1}^{\ast}-\lambda_{1}\sigma|c_{1}|^{2})}.
\\
\label{psi1_s}
\psi_{1,1}(\lambda_{1})=-\mathrm{e}^{i\lambda_{1}^{2}x}
\frac{\lambda_{1}^{\ast}\sigma
c_{1}^{\ast}}{(\lambda_{1}-\lambda_{1}^{\ast}\sigma|c_{1}|^{2})},
\\
\label{psi2_s}
\psi_{2,1}(\lambda_{1})=\mathrm{e}^{i\lambda_{1}^{2}x}
\frac{\lambda_{1}^{\ast}}{(\lambda_{1}^{\ast}-\lambda_{1}\sigma|c_{1}|^{2})}.
\end{gather}

\end{document}